\documentclass[journal,comsoc]{IEEEtran}
\usepackage[T1]{fontenc}% optional T1 font encoding

\usepackage{cite}
\usepackage{url} % for typesetting urls
\usepackage{graphicx,subfig} % Figure added by [Victoria]
\usepackage{algorithm,algorithmic} % Algorithm packages added by [Victoria]
\usepackage{amssymb} % Math font added by [Victoria]
\usepackage{pifont}  % for cross and tick marks added by [Victoria]
\usepackage{footnote}
\usepackage{tikz}  % tikz added by [Victoria]
\usepackage{adjustbox}  % tikz scale to textwidth added by [Victoria]
\usetikzlibrary{arrows,decorations.markings,positioning}  % tikz library for arrows and NN added by [Victoria]
\usepackage{multirow}  % Table added by [Victoria]
\usepackage{caption}  % Change the subfloat reference style added by [Victoria]
\usepackage{amsmath}
\usepackage{cleveref}  % added by [Victoria]
\usepackage{glossaries}
\newacronym{DNN}{DNN}{Deep Neural Network}
\newacronym{DRL}{DRL}{Deep Reinforcement Learning}
\newacronym{GAE}{GAE}{Generalized Advantage Estimation}
\newacronym{MA-DRL}{MA-DRL}{Multi-Agent Deep Reinforcement Learning}
\newacronym{MA-MDP}{MA-MDP}{Multi-Agent Markov Decision Process}
\newacronym{MA-PPO}{MA-PPO}{Multi-Agent Proximal Policy Optimization}
\newacronym{MA-RL}{MA-RL}{Multi-Agent Reinforcement Learning}
\newacronym{MDP}{MDP}{Markov Decision Process}
\newacronym{NN}{NN}{Neural Network}
\newacronym{RL}{RL}{Reinforcement Learning}
\newacronym{SDN}{SDN}{Software-Defined Networking}
\newacronym{TI}{TI}{Training Iteration}
\usepackage{enumitem} 

\ifCLASSINFOpdf
\else
\fi
\usepackage[cmintegrals]{newtxmath}
\begin{document}
\title{Multi-Agent Deep Reinforcement Learning for Request Dispatching in Distributed-Controller Software-Defined Networking}

\author{Victoria~Huang,
        Gang~Chen,
        and~Qiang~Fu% <-this % stops a space
\thanks{V. Huang is with the School of Computing \& Mathematical Sciences, University of Waikato, Hamilton 3240, New Zealand. e-mail: guiying.huang@waikato.ac.nz.
G. Chen and Q. Fu are with the School of Engineering and Computer Science, Victoria University of Wellington, Wellington 6011, New Zealand. e-mail: aaron.chen; qiang.fu@ecs.vuw.ac.nz.}}% <-this % stops a space
\maketitle

\begin{abstract}
Recently, distributed controller architectures have been quickly gaining popularity in \emph{Software-Defined Networking} (SDN). However, the use of distributed controllers introduces a new and important \emph{Request Dispatching} (RD) problem with the goal for every SDN switch to properly dispatch their requests among all controllers so as to optimize network performance. This goal can be fulfilled by designing an RD policy to guide distribution of requests at each switch. In this paper, we propose a \emph{Multi-Agent Deep Reinforcement Learning} (MA-DRL) approach to automatically design RD policies with high adaptability and performance. This is achieved through a new problem formulation in the form of a \emph{Multi-Agent Markov Decision Process} (MA-MDP), a new adaptive RD policy design and a new MA-DRL algorithm called MA-PPO. Extensive simulation studies show that our MA-DRL technique can effectively train RD policies to significantly outperform man-made policies, model-based policies, as well as RD policies learned via single-agent DRL algorithms.
\end{abstract}

\begin{IEEEkeywords}
Multi-agent deep reinforcement learning, policy gradient, request dispatching, distributed controllers, Software-Defined Networking.
\end{IEEEkeywords}

% \fi
\IEEEpeerreviewmaketitle

\section{Introduction}\label{sec:introduction}
\IEEEPARstart{A}{s} an emerging computer networking paradigm, \gls{SDN} empowers network operators with flexible network management and rapid network policy deployment~\cite{kreutz2015software}. To provide sufficient processing capacity for the increasing communication activities in a network, distributed controller architectures featuring the joint use of multiple controllers are quickly gaining popularity~\cite{onos,huang2017blac}, driving innovations to handle large operator networks such as Content Delivery Networks (CDNs)~\cite{fu2018taming}. Aiming at properly distributing its requests among all controllers so as to make the best use of controller capacity and achieve high network performance, every SDN switch often follows a \emph{request dispatching (RD) policy} to select suitable controllers to process each newly arriving request. Clearly, carefully designing such a policy is of paramount importance to the overall functioning of multi-controller SDNs~\cite{huang2017blac,huang2019csp}.

Particularly, the designed policy must satisfy three requirements:
($R_1$)~\emph{Performance effectiveness}: The policy should guide switches to properly dispatch requests to suitable controllers to minimize the average request response time. 
($R_2$)~\emph{Adaptiveness}: Note that the number of controllers in an SDN network can change to meet the varying traffic demand. Thus, the policy should perform consistently well over different numbers of controllers. 
($R_3$)~\emph{Time efficiency}: Since RD must be performed in real time with minimum delay, the policy needs to be sufficiently efficient in practice. Therefore, policies with long processing time or frequent execution (e.g., in a per-request manner) should be avoided. 

To achieve $R_1$, existing studies~\cite{gao2016traffic,wang2016dynamic,sridharan2017multiple,sridharan2017multi,huang2020scalable} constructed mathematical models to capture the correlation between the policy and the performance objective (e.g., average request response time). Although these model-driven methods can generally provide solutions with guaranteed performance, modeling the highly complicated network requires \emph{substantial domain knowledge}. Moreover, in a highly complicated distributed computing environment (e.g., a distributed controller architecture), the response time can be caused by many factors that may not be fully captured using the proposed model~\cite{li2018model}.

Alternatively, the literature has considered either manually or automatically designing policies for resource allocation~\cite{huang2017blac,nguyen2014automatic,park2018investigating,hu2017job}. Specifically, two widely-used manually designed policies in operating systems and cloud computing are weighted round-robin and first-come-first-serve~\cite{salot2013survey}. Obviously, they cannot achieve $R_1$ due to the lack of considering propagation latency and controller workload. On the other hand, Evolutionary Computation methods have been proposed to automatically design policies for standard job shop scheduling problems~\cite{nguyen2014automatic,park2018investigating}. However, these methods have \emph{high sampling costs} since data collected from previous generations cannot be reused in the next generation. Therefore, all candidate solutions in each generation need to be reevaluated in either simulated or real-world environments.

Recently, machine learning has been successfully applied to various resource management problems~\cite{mao2016resource,tesauro2006hybrid,mao2017routing,liu2018deep,cao2018machine}. Among all machine learning algorithms, we consider \gls{DRL} to be suitable for designing the RD policy for several reasons. First, \emph{no explicit mathematical model of the underlying complex environment is required}. DRL can automatically learn the policy while interacting with the unknown dynamic environment through a trial-and-error process. Second, \emph{DRL can improve current policies based on experiences/data obtained from an old policy} through a technique known as experience replay~\cite{mnih2016asynchronous}. Thus, in comparison to an EC approach, the sampling cost of training any new policies can be greatly reduced. Third, the RD problem can be naturally formulated as an \gls{MDP} (see \Cref{sec:problem-formulation} for more details).

Despite the clear advantages offered by DRL, the direct application of existing DRL value function search methods\footnote{Value function search learns the optimal value function. The value function is used to extract the optimal policy by greedily selecting the action that maximizes the long-term rewards. More details about value function search can be found in \cite{sutton2018reinforcement}}, e.g., DQN~\cite{mnih2015human}, may not be suitable. Since the learned policy is implicitly represented, making an RD decision requires extensively enumerating the entire action space to find the action with the maximal reward, violating \emph{$R_3$}. Thus, policy search~\cite{sutton2018reinforcement} which directly learns the optimal policy by searching the policy space is more appropriate. However, there are still several major issues that must be addressed.

(1) \emph{Impractical problem formulation}: It is typical to design a policy by a single learning agent supported by global network information (i.e., fully observable environment). For example, in~\cite{mao2016resource}, a centrally trained agent must learn to dispatch jobs among a large cluster of computers. Such a central approach is prone to scalability issues~\cite{zhang2018fully}. Particularly, the use of a single agent inevitably introduces extra communication delay\footnote{This is mainly because all requests from a switch must go through the central agent before they can be forwarded to controllers for processing.}. Meanwhile, obtaining timely global information over the entire SDN network can cause substantial communication overhead~\cite{wu2011novel}. 
Even though these issues can be alleviated by employing multiple co-learning agents as demonstrated in \cite{huang2019csp}, the single-agent DRL algorithm cannot cope with inter-agent interference and localized network information, resulting in poor and unpredictable network performance. 

(2) \emph{Non-adaptive and inefficient policy design}: Typically, many existing approaches~\cite{li2018model,liu2017hierarchical,li2018deep} on resource management were DQN-based policy direct search. These approaches were designed to handle problems with a discrete action space. In the RD problem, an action can be defined as assigning a set of $K$ requests to $M$ available controllers. In this case, the size of the action space is $K^M$. 
% According to existing studies~\cite{networktraffic}, the request arrival rate at a switch can be up to 10000 pkt/s in a data center. Therefore, the action space can be large. 
With such a large action space, the policy complexity inevitably increases, resulting in long computational time to output an action. To reduce the action space, an action can also be defined as assigning one request to a controller every time where the size of the action space is $M$. However, this action definition requires the policy to be processed repeatedly with respect to every new request, incurring non-negligible policy processing overhead. Therefore, both discrete action definitions cannot satisfy \emph{$R_3$}.
Moreover, existing DRL approaches directly represent their policy as a \gls{DNN} with a fixed number of output nodes. The number of output nodes is the size of the action space. Such a representation apparently violates \emph{$R_2$} since the policy fails to function well whenever the number of controllers is changed to meet the varying traffic demand. 

(3) \emph{Inapplicable adaptive policy training}: 
% This paper aims to design an adaptive policy that can support a changing number of controllers. 
To perform policy training using policy search, it is critical to calculate the gradient of a policy~\cite{sutton2018reinforcement}. In existing DRL (e.g., PPO~\cite{schulman2017proximal}, TRPO~\cite{trpo}), the gradient of a policy is essentially the gradient of a \gls{DNN}, which is straightforward. However, a policy that can support a changing number of controllers obviously cannot be directly represented as a \gls{DNN}. In this case, how to compute its gradient needs to be addressed.
% However, the requirement of training an adaptive policy cannot be satisfied by directly following existing DRL-based approaches~\cite{mao2016resource,tesauro2006hybrid}. In standard DRL (e.g., PPO~\cite{schulman2017proximal}, TRPO~\cite{trpo}), calculating the gradient of a \gls{DNN} policy with a fixed number of output nodes is straightforward. However, when a policy that can support a changing number of controllers, how to compute its gradient needs to be addressed.

As far as we know, none of the existing studies have considered and solved the above issues. In this paper, we develop a new \emph{Multi-Agent DRL}~(MA-DRL) approach for learning adaptive policies for SDN switches. Our research successfully addresses the three limitations above with the key novelties as summarized below:

A.~\emph{Practical formulation of the policy design problem as a Multi-Agent Markov Decision Process~(MA-MDP)}: Rather than using a centralized agent, we equip each distributed switch with a co-located agent. All agents share the same goal of optimizing the network-wide communication performance in terms of the average request response time. Without assuming fully observable agents supported by global network knowledge, a partially observable network is considered where each agent executes its policy based only on its local observation.

B.~\emph{Adaptive policy design for efficient request dispatching over an arbitrary number of controllers}: We propose a new DNN-based policy representation to allow any switch to distribute its requests among a changing number of controllers. To satisfy $R_3$, our policy outputs the controller priorities which are mapped to probabilities to guide the RD among eligible controllers within a specific time period. The priority output is hence treated as the continuous action in the context of MA-DRL. Guided by such actions, each request can be quickly dispatched in a probabilistic manner without repeated processing of the policy network. Meanwhile, we embed a controller filtering mechanism into our dispatching system to prevent controller overloading as well as unnecessary use of distant controllers. 
% This mechanism ensures health network operation during online training of the policy.

C.~\emph{Multi-agent training of an adaptive policy}: To support the training of adaptive policies, a new mathematical technique is developed to calculate the policy gradient. Meanwhile, we develop a policy training system to fulfill the general principle of centralized training and decentralized execution \cite{maddpg}, which is essential for reliable MA-DRL. However, using the Deep Deterministic Policy Gradient (DDPG) algorithm as recommended in~\cite{maddpg} is not suitable for our problem because it needs to learn the centralized Q-function with large function input that contains the multi-agent joint action space. Instead, we implement a new multi-agent version of the Proximal Policy Optimization (PPO) algorithm~\cite{schulman2017proximal} that only requires to learn the centralized value function with substantially reduced input dimensions.

% In summary, we propose an MA-DRL approach to address the policy design problem for high network performance. Specifically, the problem is first formulated as an MA-MDP in a partially observable content. Along with the formulation, an adaptive policy is designed which generates the priorities over an arbitrary number of controllers. To support the training of such an adaptive policy, a training system is developed armed with a new gradient calculation technique which enables reliable MA-DRL.

\section{Related Work}\label{sec:related-work}
Generally speaking, the RD problem in SDN can be solved at two granularity levels: switch level and request level. 
% Let us consider a network with $N$ switches and $M$ controllers. $K$ requests generated by $N$ switches within a specific period need to be dispatched to $M$ controllers. 
When RD is performed at the switch level, solving the RD problem means finding the switch-controller mapping.
% which has a complexity of $\mathcal{O}(M^N)$. 
On the contrary, when solving the RD problem at the request level, requests from one switch are no longer restricted to be handled by only one controller. Instead, they can be flexibly distributed and processed among multiple controllers. However, this flexibility comes at a cost of increasing the problem complexity which will be discussed in \Cref{subsec:request_level_rd}.
% from $\mathcal{O}(M^N)$ to $\mathcal{O}(M^K)$ given $N \ll K$. Although solving the request dispatching problem at the switch level seems more feasible, it is still NP-hard according to existing studies~\cite{wang2017switch,wang2016dynamic,wang2017efficient,cello2017balcon}. 

\subsection{Switch-level Request Dispatching}
Different approaches were proposed to find the switch-controller mapping, e.g., approximation algorithms~\cite{gao2016traffic,cheng2015dha,cheng2015qos,ye2017maximizing} and heuristics~\cite{wang2017switch,cheng2015qos,elasticon,cui2018load,xu2019dynamic}. 
% and game theory~\cite{wang2016dynamic,chen2016game,cheng2016dynamic}. 

Gao et al.~\cite{gao2016traffic} formulated the RD problem as an integer programming problem with the goal of balancing the workload among controllers. The problem was transferred into linear programming using relaxation and solved by an approximation algorithm called deterministic rounding. Related approximation approaches can be found in~\cite{cheng2015qos,cheng2015dha,ye2017maximizing}.

Due to the NP-hardness of the switch-controller mapping problem~\cite{wang2017switch,wang2016dynamic}, it is computationally expensive to find the optimal solution in a large network. Thus, heuristic methods have been widely used~\cite{wang2017switch,cheng2015qos,elasticon,cui2018load,liang2014scalable}. In~\cite{liang2014scalable}, whenever the load difference between the heaviest-loaded and lightest-loaded controllers is greater than a predefined threshold, a switch will be mapped from the heaviest-loaded controller to the lightest-loaded one. Similar greedy approaches have also been used in~\cite{wang2017switch,bari2013dynamic,cui2018load}. Although these heuristic approaches can find an acceptable solution of the problem within a reasonable time, they cannot guarantee the solution quality.  

% On the other hand, game theory has also been used to solve the switch-controller mapping problem. For example, 
% % Wang et al.~\cite{wang2016dynamic} first transformed the switch-controller mapping problem into a classical college admissions problem and solved using matching theory to guarantee the worst-case performance. Then the obtained switch-controller mapping was fed into the coalitional game where switches were transferred between coalitions (i.e., controllers) to achieve a Nash stable solution. 
% a zero-sum game based approach was proposed in~\cite{chen2016game}. Specifically, a game was initialized with a randomly selected switch (i.e., a commodity) managed by an overloaded controller. All neighboring available controllers were invited as game players for competition from which a new destination controller was selected with maximal utility changes. Other game theory based approaches can be found in~\cite{cheng2016dynamic,wang2016dynamic}. 

Note that all algorithms in this category assume that requests generated from one switch can only be handled by its mapped controller. Whenever the switch is re-mapped from one controller to another, the switch's workload will be all transferred to the new controller, rendering the new controller susceptible to being overloaded. Moreover, the RD can only be performed at a coarse level (i.e., switch level), restricting the opportunity of properly distributing workload across all controllers to achieve high network performance, which has been demonstrated in~\cite{sridharan2017multiple,huang2017blac}. 

\subsection{Request-level Request Dispatching}\label{subsec:request_level_rd}
To address the issues of switch-level RD methods, approaches designed for request-level RD have been proposed. In this category, the requests from one switch can be distributed and processed among multiple controllers.

For example, BalanceFlow~\cite{balanceflow} employed a super controller which ran a greedy heuristic to partition the overall workload of the control plane among all controllers based on the collected global traffic information. However, due to its dependence on the super controller, BalanceFlow may not scale well in large networks. To address this issue, BLAC~\cite{huang2017blac} introduced a scheduling layer where multiple schedulers were deployed to distribute requests from switches to different controllers. Similarly, to reduce the overhead caused by switch-controller remapping~\cite{elasticon} and balance the controller workload, Al-Tam et al.~\cite{al2019fractional} partially transferred the workload from an overloaded controller to other underloaded controllers.
 % periodically collect the global traffic information from the network. Then the super controller ran a greedy heuristic algorithm to partition the workload among all controllers. 
% The partition decision was transformed as flow rules installed in the switches. Thus, packets arriving at the switch matching these rules will be directly forwarded to corresponding (different) controllers, reducing the workload of the original controller. 
Similar work can also be found in~\cite{selvi2016cooperative,yao2015multicontroller,sridharan2017multi,huang2020scalable}. 

To capture the correlation between the RD policy and its performance, a queuing model was adopted in~\cite{sridharan2017multi} and an improved round-robin heuristic was proposed to determine the request distribution among switches and controllers. Similarly,~\cite{huang2020scalable} formulated the RD problem as an optimization problem with the goal of minimizing the average response time. A Gradient-Descent-based (GD) algorithm was developed to calculate the suitable request distribution.

Compared to the switch-level RD, request-level RD enables request dispatching to be performed at a fined-grained level. The benefits of request-level RD have also been demonstrated in~\cite{huang2017blac,sridharan2017multi,chattopadhyay2019aloe}. Therefore, this paper will focus on request-level RD. However, the flexibility of request-level RD comes at a cost of increasing the problem complexity from $\mathcal{O}(M^N)$ to $\mathcal{O}(M^K)$ where $M,N,K$ are the number of controllers, switches, and requests with~$N \ll K$. The significantly increased problem complexity rendering the efficiency and effectiveness of the switch-level RD methods questionable. 
Although different request-level RD approaches have been proposed, existing works have certain limitations. For example, the introduction of a centralized super controller limits the scalability of the control plane~\cite{yao2015multicontroller,sridharan2017multiple}. Apart from that, existing works mostly used heuristics which cannot guarantee the quality of the solution (see \Cref{subsubsec:ma_simulation}). Although a GD-based algorithm was proposed in~\cite{huang2020scalable}, the network performance achieved by GD relies on the provided network information accuracy. But network information such as request arrival rates can only be estimated in practice. The inaccurate information hinders GD achieving its optimal performance, which will be demonstrated in \Cref{subsubsec:ma_simulation}. 

\subsection{DRL-based Resource Management}
Recently, \gls{DRL} has been successfully utilized in many resource management problem~\cite{tesauro2006hybrid,mao2016resource,mao2019learning,nazari2018reinforcement,chinchali2018cellular,gu2020mddpg,roig2020mano,wu2019nfc,zhang2020CFR-RL,li2018deep}.
 % since resource management can be naturally mapped to a sequential decision making process.

For example, Li et al.~\cite{li2018deep} applied DRL to address the 5G network slicing problem. Specifically, given a fixed number of existing slices with the shared aggregated bandwidth and the demands of each slice, the agent trained by DQN dynamically adjusts the bandwidth sharing to maximize the resource utilization while maintaining high user experience satisfaction. Similarly, Hua et al.~\cite{hua2019gan} proposed a generative adversarial network-powered deep distributional Q-network to allocate network resources for diversified services in 5G networks. 
Moreover, Tesauro et al.~\cite{tesauro2006hybrid} proposed an RL-based approach to automatically allocate the server resources in data centers. DeepRM~\cite{mao2016resource} was proposed to address the multi-resource cluster scheduling problem using a \gls{DNN} policy to optimize various objectives, e.g., average job completion time and resource utilization. Similarly, Decima~\cite{mao2019learning} combined DRL and graph neural networks to learn workload-specific scheduling policies for data processing clusters. Chinchali et al.~\cite{chinchali2018cellular} leveraged the delay-tolerant feature of IoT traffic and developed an RL-based scheduler to handle traffic variation so that the network utilization can be constantly optimized. 

Different from heuristics, DRL fully automates the policy design process and noticeably improves the performance of designed policies~\cite{mao2019learning}. However, many existing approaches are designed under the assumption of a single agent and fully observable environment which cannot be easily satisfied in our RD problem as we mentioned in \Cref{sec:introduction}. 

To address the above issues, MA-DRL techniques have been developed in \cite{yang2019large,qiu2019dynamic}. In particular, \cite{ye2015multi} proposed a multi-agent Q-learning system to guide packet routing in wireless sensor networks. Similar studies can also be found in~\cite{wu2018decentralised,noureddine2017multi,wu2018decentralised}. Despite the promising progress, these studies rely heavily on inter-agent communication, which may introduce non-negligible communication overhead, unsuitable for RD in SDNs. Furthermore, most of the policies trained via MA-DRL cannot adaptively support a varying number of actions and thus may not scale well to large networks~\cite{abdallah2006learning,abdallah2007multiagent}.

\section{The Policy Design Problem in SDN}\label{sec:problem-formulation}
\begin{figure}[]
\centering\includegraphics[width=.85\linewidth]{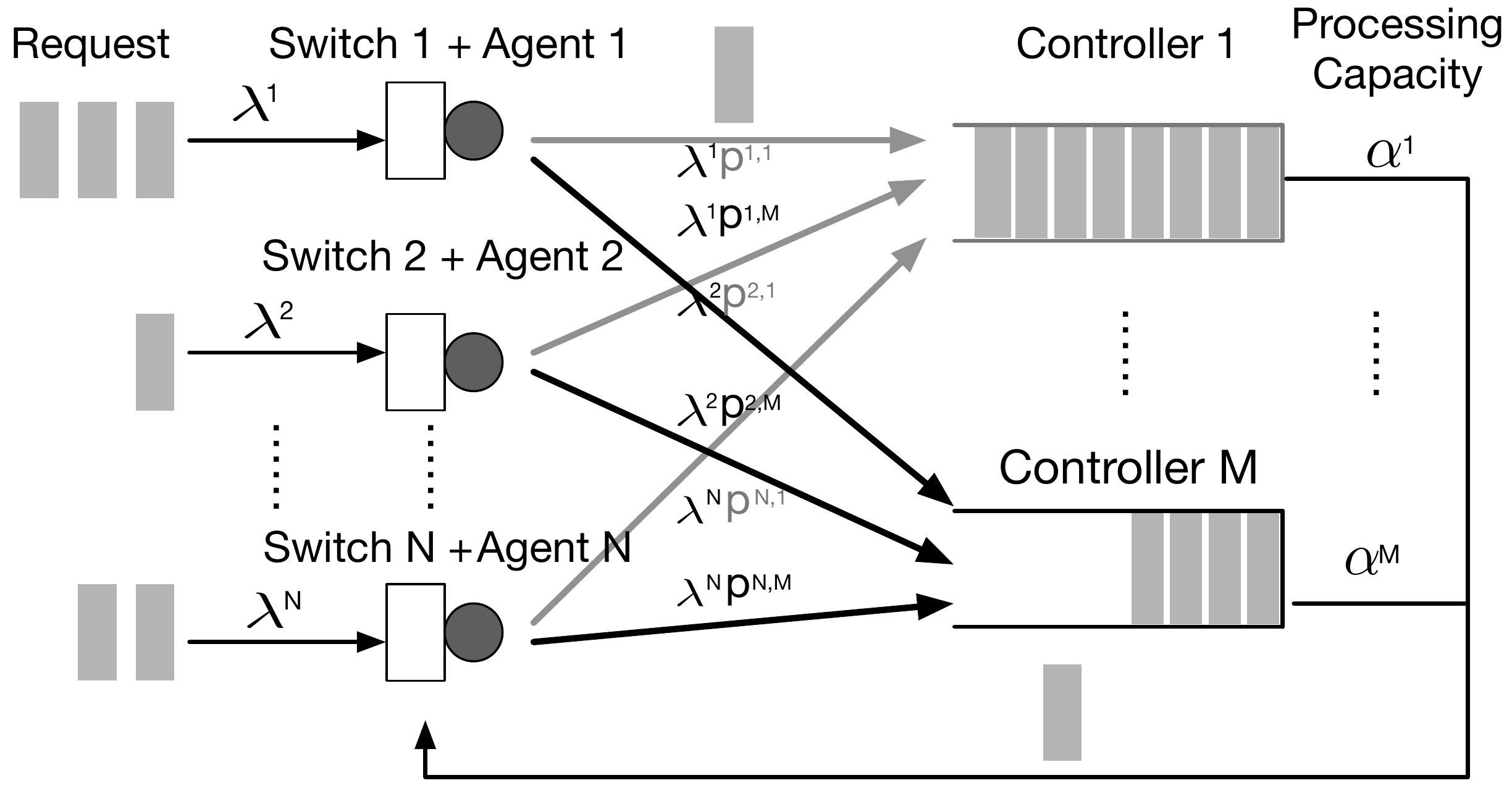}
\vspace{-5pt}
\caption{Request Dispatching in SDN.}
\label{fig:request_dispatching}
\end{figure}

Figure~\ref{fig:request_dispatching} shows an SDN network with $N$ switches and $M$ controllers. The processing capacities of these controllers are determined as $\boldsymbol \alpha = [\alpha^1, ..., \alpha^{M}]$. The one-way propagation latency between switches and controllers is measured in matrix $\boldsymbol D$ where each element $D_{m,n}$ refers to the propagation latency between switch $Sw_n$ and controller $C_m$. Whenever a new packet arrives at a switch, the switch will generate a request and forward it to a controller chosen by the agent for processing. The request generation rate at every switch is denoted as $\boldsymbol \lambda = [\lambda^1,..., \lambda^{N}]$. Each controller processes its requests in a FIFO manner~\cite{huang2017blac}. After processing a request, the corresponding response will be sent back from the controller to the switch. The time interval measured by the switch from sending a request to receiving a response is denoted as the \emph{request response time} $\tau$.

In this paper, the policy design problem is modeled as a \emph{fully cooperative} and \emph{partially observable} MA-MDP with $N$ agents that control the $N$ SDN switches respectively, as shown in \Cref{fig:request_dispatching}. The overall network operating status is captured by a set of global states $\mathcal{S}$. At each time step~$t$, the network enters a state $s_t \in \mathcal{S}$.
Every agent $Agt_n$ receives a local observation $z^{n,m}_t$ with respect to each controller $C_{m}$. The relationship between $z^{n,m}_t$ and $s_t$ is determined by the agent's observation function $z^{n,m}_t=\mathcal{O}^n(s_t,m)$. Based on the local observations, every agent $Agt_n$ issues an action $\boldsymbol a^n_t \in \mathcal{A}^n$ chosen from its policy $\pi_{\boldsymbol \theta^n}$ to jointly form the multi-agent action $\{\boldsymbol a_t^n\}_{n=1}^N$. Here $\boldsymbol a^n_t=\{a^{n,m}_t\}_{m=1}^M$ specifies the priority $a^{n,m}_t$ for $Agt_n$ to dispatch its new requests to any controller $C_m$ during time $t$ and $t+1$. As a result of following the joint action, each agent obtains a reward $r^n_t$ based on the responses it received from all controllers during time $t$ and $t+1$, as defined below:
\begin{equation}
  \label{eqt:agent_individual_reward}
  r^n_t = \varsigma X^n_t - \sum_{x=1}^{X^n_t} \tau^n_x,
\end{equation}
where $X^n_t$ is the total number of responses received between $t$ and $t+1$ by $Agt_n$ and $\tau^n_x$ is the response time of a particular request. $\varsigma$ is a \emph{weight factor} that controls the importance of the throughput~$\chi_t^n$ relative to the response time. In our simulation, $\varsigma$ is estimated as the average response time of weighted round robin~\cite{huang2020scalable} which serves as a baseline for the policy. Clearly, all agents prefer to receive more responses with lower request response time according to~\eqref{eqt:agent_individual_reward}. For this purpose, $Agt_n$ learns a policy $\pi_{\boldsymbol \theta^n}$ parameterized by $\boldsymbol \theta^n$ that maps observation $\boldsymbol z_t^n = \{z_t^{n,m}\}_{m=1}^M$ to its action $\boldsymbol a_t^n$. More details on the adaptive policy design will be presented in \Cref{subsec:policy-design}.
% with the adaptive design shown in \Cref{fig:cha5_sa_policy} that maps observation $\boldsymbol z_t^n = \{z_t^{n,m}\}_{m=1}^M$ to its action $\boldsymbol a_t^n$. 
The goal of MA-MDP is hence to identify the optimal policies $\{\pi^*_{\boldsymbol \theta^n}\}_{n=1}^N$ so as to maximize the \emph{expected joint cumulative rewards}:
\begin{equation}
  \label{eqt:madrl_objective}
  J(\{\pi_{\boldsymbol \theta^n}\}_{n=1}^N) =
  \mathbb{E}_{\{\boldsymbol a_t^n \sim \pi_{\boldsymbol \theta^n}\}_{n=1}^N} \sum_{t=0}^{T} \gamma^t \sum_{v \in V} r^n_t(\boldsymbol z_t^{n}, \boldsymbol a^n_t)
\end{equation} 
where $\gamma \in [0,1)$ is a discount factor. Evaluation with different $\gamma$ values will be reported in \Cref{subsubsec:hist_gamma_impact}.

\section{MA-DRL for RD Policy Learning} 
In line with the MA-MDP problem defined in \Cref{sec:problem-formulation}, an adaptive DNN-based policy will be proposed in this section to ensure efficient RD over an arbitrary number of controllers. A training system will be subsequently developed to train the adaptive RD policy with a newly developed mathematical technique to estimate the policy gradient.

\subsection{DNN-based Adaptive Policy Design}\label{subsec:policy-design}
An RD policy is expected to adapt easily to a changing number of controllers. However, this requirement is seldom supported by existing policy representations which only allow a fixed collection of actions as we mentioned in \Cref{sec:introduction}. 
One possible strategy to solve this issue is to train multiple policies while each policy targeting at a particular number of controllers. However, 
% the cost of evaluation or training a policy in a production network can be high. This is mainly because production networks should always guarantee reasonable good performance while the policy can perform badly (e.g., overloading some controllers which leads to long response time) especially at the early training stage. Furthermore, 
each policy needs to be individually evaluated or trained in advance before being deployed, which leads to high sampling costs. Thus, instead of training multiple policies, we should design and train an adaptive policy that can support different numbers of controllers.

  \pgfdeclarelayer{background}
  \pgfsetlayers{background,main}

  \begin {figure*}%[!hbtp]
  \centering
  \begin{adjustbox}{width=.85\linewidth}
  \begin{tikzpicture}[scale=.4,cap=round]
      
      % draw arrow from observation to extraction
      \draw [->] (-7, 3.2) -- (-4, 3.2);
      \node [align=center, above, font=\small] at (-5.5, 3.2) { $s_t$};

      \node [align=center, above, font=\footnotesize] at (-6, 6.5) { Global\\ state};

      % draw the rectangle for state extraction
      \draw [fill=white] (-4,2.5) rectangle (-1,4.5);
      \node [font=\small] at (-2.5, 3.5) {$\mathcal{O}^n$};

      % draw arrows from state extraction to NN and write notations
      \draw [->] (-1, 3.2) -- (1.7, 3.2); 
      \node [font=\small] at (0.3, 3.7) {$z_{t}^{n,m}$};
      
      \draw [->] (-1, 4.5) to [out=45, in=135] (1.7, 4.5);
      \node [font=\small] at (0.3, 5.8) {$z_{t}^{n,1}$};

      \draw [->] (-1, 2.5) to [out=-45, in=-135] (1.7, 2.5);
      \node [font=\small] at (0.3, 1) {$z_{t}^{n,M}$};

      \node [align=center, above, font=\footnotesize] at (0.0, 6.5) { Controller\\ state};

      % draw vertical dots
      \foreach \x in {0.3}
          \foreach \y in {4.5, 4.7, 4.9, 2.2, 2.4, 2.6 } {
            \filldraw[fill=black] (\x,\y) circle (0.03);
          }

      % draw the rectangle around the NN
      \draw (1.7, 0.5) -- (1.7, 6.3) -- (9.3, 6.3) -- (9.3, 0.5) -- (1.7, 0.5);
      \node [font=\small] at (8.5, 1.5) {$f_{\boldsymbol\theta^n}$};

      \node [align=center, above, font=\footnotesize] at (5.7, 6.3) { Priority function \\ (DNN)};

      % === draw NN
      % draw the nodes
      \foreach \x in {2}
          \foreach \y in {2,3,4,5} {
          \filldraw[fill=white] (\x,\y) circle (0.1);
          }
      \foreach \x in {9}
          \foreach \y in {3.5} {
          \filldraw[fill=white] (\x,\y) circle (0.1);
          }
      \foreach \x in {4,7}
          \foreach \y in {1,2,3,4,5,6} {
              \filldraw[fill=white] (\x,\y) circle (0.1);
          }
         
      \begin{pgfonlayer}{background}
          % we add the lines for the nodes starting in y 2,3, and 4
          \foreach \xa / \xb in {2 / 4, 4 / 7 , 7 / 4}
              \foreach \ya / \yb / \yc / \yd / \ye / \yf in {2 / 3 / 4 / 5 / 1 / 6, 3 / 4 / 5 / 1 / 2 / 6, 4 / 5 / 1 / 2 / 3 / 6, 5 / 3 / 4 / 5 / 1 / 6} {
                  \draw (\xa,\ya) -- (\xb,\ya);
                  \draw (\xa,\ya) -- (\xb,\yb);
                  \draw (\xa,\ya) -- (\xb,\yc);
                  \draw (\xa,\ya) -- (\xb,\yd);
                  \draw (\xa,\ya) -- (\xb,\ye);
                  \draw (\xa,\ya) -- (\xb,\yf);
              }
          % add remaining lines from y1 to y5
          \foreach \xa / \xb in {4 / 7 , 7 / 4}
              \foreach \ya / \yb in {1 / 5, 5 / 1} {
              \draw (\xa,\ya) -- (\xb,\ya);
              \draw (\xa,\ya) -- (\xb,\yb);
          }
          % add lines from x11 to x16
          \foreach \xa / \xb in {7 / 9}
              \foreach \ya / \yb in {1 / 3.5, 2 / 3.5, 3 / 3.5, 4 / 3.5, 5 / 3.5, 6 / 3.5}{
              \draw (\xa,\ya) -- (\xb,\yb);
          }
      \end{pgfonlayer}

      % draw arrows from NN to softmax and write notations
      \draw [->] (9.3, 3.2) -- (13, 3.2); 
      \node [font=\small] at (11, 3.7) {$o_{t}^{n,m}$};
      
      \draw [->] (9.3, 4.5) to [out=45, in=135] (13, 4.5);
      \node [font=\small] at (11, 6) {$o_{t}^{n,1}$};

      \draw [->] (9.3, 2.5) to [out=-45, in=-135] (13, 2.5);
      \node [font=\small] at (11, 1) {$o_{t}^{n,M}$};

      \node [align=center, above, font=\footnotesize] at (10.8, 6.3) { Controller \\ priorities};

      % draw vertical dots
      \foreach \x in {11}
          \foreach \y in {4.5, 4.7, 4.9, 2.2, 2.4, 2.6 } {
            \filldraw[fill=black] (\x,\y) circle (0.03);
          }

      % softmax mapping
      \draw [fill=white] (13,2.5) rectangle (16,4.5);
      \node [font=\scriptsize] at (14.5, 3.5) {Softmax};
      
      % arrow from softmax to noise 
      \draw [->] (16, 3.2) -- (20, 3.2); 
      \node [font=\small] at (18, 3.7) {$\tilde o_{t}^{n,m}$};
      
      \draw [->] (16, 4.5) to [out=45, in=135] (20, 4.5);
      \node [font=\small] at (18, 6) {$\tilde o_{t}^{n,1}$};

      \draw [->] (16, 2.5) to [out=-45, in=-135] (20, 2.5);
      \node [font=\small] at (18, 1) {$\tilde o_{t}^{n,M}$};

      % draw vertical dots
      \foreach \x in {18}
          \foreach \y in {4.5, 4.7, 4.9, 2.2, 2.4, 2.6 } {
            \filldraw[fill=black] (\x,\y) circle (0.03);
          }

      % Adding sum rectangle
      \draw [fill=white] (20,2.5) rectangle (23,4.5);
      \node [font=\small] at (21.5, 3.5) {$\sum$};

      % noise arrow
      \draw [->] (21.5, 5) -- (21.5, 4.5);
      \node [align=center, above, font=\scriptsize] at (21.5, 5) {Noise\\$\boldsymbol{\epsilon}^n_t \sim \mathcal{N}(0, \boldsymbol{\Sigma})$};

      % arrow from sum to controller filtering
      \draw [->] (23, 3.2) -- (29, 3.2); 
      \node [font=\small] at (26.5, 4) {$\{a_{t}^{n,m}\}_{m=1}^M$};
      
      \node [align=center, above, font=\footnotesize] at (26.5, 6) {Action};

      % Draw policy rectangle
      \draw [fill=none, dashed] (1.5, 0) rectangle (24, 8.5); 
      \node [align=center, above] at (22, 0) {\textbf{Policy}};
  \end{tikzpicture}
  \end{adjustbox}
  % \vspace{-5pt}
  \caption{The DNN-based Adaptive Policy Design.}
  \label{fig:cha5_sa_policy}
  \end{figure*}
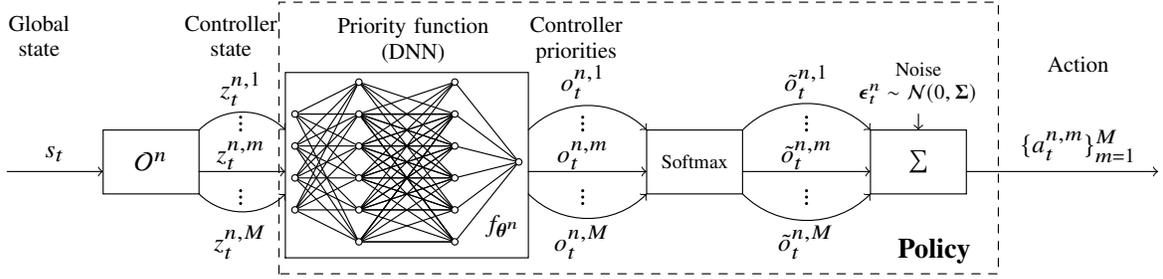

To address this issue, we propose a new policy design, as shown in \Cref{fig:cha5_sa_policy}. In particular, the policy $\pi_{\boldsymbol \theta^n}$ takes the local observations $\boldsymbol z_t^n$ from agent $Agt_n$ as inputs and outputs an action $\boldsymbol a_t^n$. In our previous work~\cite{huang2019csp}, an action corresponds to a chosen controller for request processing. This design requires repeated processing of the policy with respect to every new request, preventing efficient use of the policy in traffic-intensive networks. This issue is addressed by defining $\boldsymbol a_t^n=\{a^{n,m}_t\}_{m=1}^M$ as the controller priorities to guide the request dispatching, which is updated once in every given time interval as discussed earlier.

\emph{Priority mapping}: 
Instead of generating the action $\boldsymbol a_t^n=\{a^{n,m}_t\}_{m=1}^M$ through one run of the policy network, an agent $Agt_n$ feeds its local observation $z_t^{n,m}$ with respect to controller $C_m$ to the DNN in \Cref{fig:cha5_sa_policy} one-by-one for all controllers.
% As we discussed in \Cref{sec:introduction}, existing policies are generally represented as a \gls{DNN} (e.g., policies trained using PPO~\cite{schulman2017proximal}). These policies generate the request dispatching probabilities $\{p^{n,m}_t\}_{m=1}^M$ through one run of the DNN, which cannot adapt to a changing number of controllers. In our new policy design, the specific state information $z_t^{n,m}$ with respect to each controller $C_m$ is first extracted from $\boldsymbol z_t^{n}$. Then $z_t^{n,m}$ is fed one-by-one to the DNN in \Cref{fig:cha5_sa_policy} for all controllers. 
For each local observation $z_t^{n,m}$, the DNN assigns a priority value $o_t^{n,m}$ to $C_{m}$. For simplicity, we denote the DNN as a priority function $f_{\boldsymbol\theta^n}$ with trainable parameters $\boldsymbol \theta^n$, to distinguish it from the policy $\pi_{\boldsymbol \theta^n}$ with additional components for \emph{normalization} and \emph{exploration}\footnote{The exploration component in a policy is only activated during policy training for stochastic exploration of different request dispatching distributions. While testing the trained policy on a SDN network, this component is deactivated.}, as explained below.
  
\emph{Normalization and exploration}: The softmax function is used to normalize all controllers' priorities $\{o_t^{n,m}\}_{m=1}^M$ into a probability distribution $\{ \tilde{o}_t^{n,m}\}_{m=1}^M$, as indicated in \Cref{fig:cha5_sa_policy}. Rather than using $\{ \tilde{o}_t^{n,m}\}_{m=1}^M$ in a deterministic manner, the agent must continue to explore different request dispatching distributions and determine their impact on network performance during policy training. This is achieved by adding small Gaussian noises
    \begin{equation*}
    \epsilon_t^{n,m} \sim \mathcal{N}(0, \sigma^2), \qquad m=1,..,M
    \end{equation*}
    to $\tilde{o}_t^{n,m}$, as defined below:
    \begin{equation}
        \boldsymbol a_t^n = \boldsymbol{\tilde{o}}_t^n + \boldsymbol \epsilon_t^n,
    \label{eqt:rho_sa}
    \end{equation}
where $\boldsymbol{\tilde{o}}_t^n = \{\tilde{o}_t^{n,m}\}_{m=1}^M$ and $\boldsymbol \epsilon_t^n=\{\epsilon_t^{n,m}\}_{m=1}^M$. 

In association with the discussion above, the whole action generation process based on our new policy design can be formulated as:
    \begin{equation}
    \resizebox{1.\hsize}{!}{$
       \boldsymbol{a}_t^n= \begin{bmatrix} a^{n,1}_t\\ \vdots\\ a^{n,M}_t\end{bmatrix}
        =\pi_{\boldsymbol\theta^n} \left (  \begin{bmatrix} z^{n,1}_t\\ \vdots\\ z^{n,M}_t\end{bmatrix} \right)
        = \begin{bmatrix} \text{Softmax}(f_{\boldsymbol{\theta}^n}(z^{n,1}_t)) +\epsilon_t^{n,1}  \\ \vdots\\ \text{Softmax}(f_{\boldsymbol{\theta}^n}(z^{n,M}_t)) +\epsilon_t^{n,M} \end{bmatrix}
    \label{eqt:pi_design}
    $}
    \end{equation}
Because of $\boldsymbol \epsilon_t^n$, \eqref{eqt:pi_design} produces $\boldsymbol{a}_t^n$ as the continuous action output in a stochastic manner. 

% \begin{figure}[!tbp]
% \centering\includegraphics[width=0.8\linewidth]{figures/chapter5/ma_mdp.pdf}
% \caption{Modeling the policy design problem as an MA-MDP.}
% \label{fig:chap5_ma_mdp}
% \end{figure}

\subsection{The Dispatching System Design}\label{subsec:dispatching_system_design}
  
% ============ mapping from priorities to RD probabilities
  \pgfdeclarelayer{background}
  \pgfsetlayers{background,main}

  \begin {figure*}%[!hbtp]
  \centering
  \begin{adjustbox}{width=.85\linewidth}
  \begin{tikzpicture}[scale=.35,cap=round]
    % state observation
      \node [align=center, above, font=\footnotesize] at (15, 5) {Local\\Observation};

    % arrow from observation to policy
      \draw [->] (12, 3.2) -- (18, 3.2); 
      \node [font=\small] at (15, 4) {$\{z_{t}^{n,m}\}_{m=1}^M$};
      
    % RD policy
      \draw [fill=white] (18,2.5) rectangle (22,4.5);
      \node [align=center, above, font=\scriptsize] at (20, 2.8) {policy};

      % action
      \node [align=center, above, font=\footnotesize] at (24, 4.5) {Action};

      % arrow from sum to controller filtering
      \draw [->] (22, 3.2) -- (27, 3.2); 
      \node [font=\small] at (24.5, 4) {$\{a_{t}^{n,m}\}_{m=1}^M$};
      
      % controller filtering
      \draw [fill=white] (27,2.5) rectangle (31,4.5);
      \node [align=center, above, font=\scriptsize] at (29, 2.5) {Controller\\filtering};

      % controller candidate
      \draw [->] (29, 5) -- (29, 4.5);
      \node [align=center, above, font=\scriptsize] at (29, 5) {Candidate Controller List\\ $\{L_t^{n,m}\}_{m=1}^M$};

      % arrow from controller filtering to probability mapping
      \draw [->] (31, 3.2) -- (36, 3.2); 
      \node [font=\small] at (33.5, 4) {$\{\tilde a_{t}^{n,m}\}_{m=1}^M$};

      % Probability mapping
      \draw [fill=white] (36,2.5) rectangle (40,4.5);
      \node [align=center, above, font=\scriptsize] at (38, 2.5) {Probability\\mapping $T$};

      \draw [->] (40, 3.5) -- (48, 3.5);
      \node [font=\footnotesize, above] at (44, 3.5) {$\{ p_{t}^{n,m}\}_{m=1}^M$};

      \node [align=center, above, font=\footnotesize] at (45, 5) {Request Dispatching \\Probabilities};

      % Draw policy rectangle
      \draw [fill=none, dashed] (17.5, 0.7) rectangle (41, 7); 
      \node [align=center, above] at (35, 0.5) {\textbf{Dispatching System}};
  \end{tikzpicture}
  \end{adjustbox}
  % \vspace{-5pt}
  \caption{The design of the dispatching system.}
  \label{fig:cha5_ac2prob}
  \end{figure*}
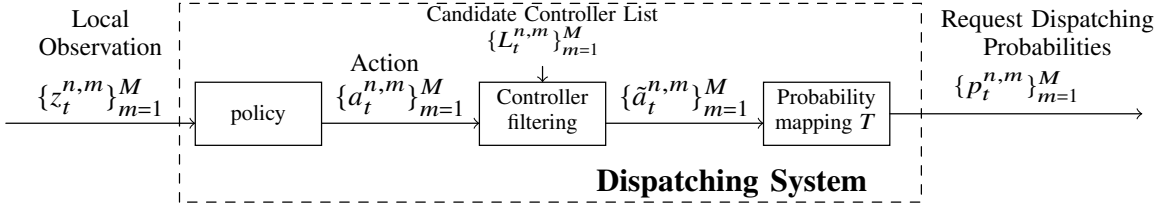
 
  When performing request dispatching, the switch should avoid sending requests to unsuitable controllers, e.g., overloaded or remotely located controllers. Driven by this motivation, a \emph{controller filtering mechanism} is designed and used before mapping $\boldsymbol{a}_t^n$ to RD probabilities. In particular, the agent keeps track of the operating status of all controllers\footnote{Controller status update is realized through regular beacon messages send by every controller to the agent in the network. Due to the communication overhead, beacon messages are not communicated at high frequencies. Hence, the status information accessible to the agent can be slightly outdated. Despite this, simulation studies in \Cref{subsec:simulation_result} show that the agent can achieve high network performance via DRL.} and maintains a candidate controller list $\boldsymbol{L}^n_t = \{L_t^{n,m}\}_{m=1}^M$. Preference is given to controllers with relatively small propagation latency from the agent as well as controllers under moderate or low workload\footnote{The average queue length of a controller must fall below a predefined threshold for the controller to be considered for request processing.}. Accordingly, up to $\chi$ controllers can be considered as candidates by the agent.
 % where $\chi$ can be flexibly set to control the level of exploration. For example, when $\chi=N$, all controllers are considered for request dispatching. Similarly, the agent can only explore at most $\chi$ controllers when $\chi<N$.

  Armed with the adaptive policy and the controller filtering mechanism, a dispatching system is designed for each switch as shown in \Cref{fig:cha5_ac2prob}. The dispatching system takes the local observations as inputs and outputs the request dispatching probabilities for the switch. In particular, given the local observations $\{z_{t}^{n,m}\}_{m=1}^M$, the policy generates the action $\{a_{t}^{n,m}\}_{m=1}^M$ according to~\eqref{eqt:pi_design}. After that, overloaded or remotely located controllers are filtered by assigning 0 to the corresponding $a_t^{n,m}$ with the help of $\{L_{t}^{n,m}\}_{m=1}^M$. The filtered action $\{\tilde a_{t}^{n,m}\}_{m=1}^M$ is then mapped to $\{p^{n,m}_t\}_{m=1}^M$ through function $T$. Mathematically, $\boldsymbol{L}_t^n$ is presented as a binary vector that covers all the $N$ controllers, with the corresponding elements of $\boldsymbol{L}_t^n$ for candidate controllers taking the value 1.

\subsection{MA-PPO for Adaptive Policy Training}\label{subsec:ma_ppo}
Aiming at training an policy for each SDN switch in a network, one straightforward approach is to directly adopt the single-agent \gls{DRL} algorithm. In particular, one DRL agent is placed on every SDN switch to continuously and independently learn its policy while the other agents are treated as part of the environment. Despite its simplicity, the training process is vulnerable to the \emph{non-stationary} environment problem~\cite{maddpg}. In particular, the reward received by each agent and the global state transition do not depend solely on one agent's individual actions. Instead, they are affected by the joint actions from all agents. Moreover, each agent's policy keeps being updated independently during the training process. Therefore, the environment observed by each agent becomes non-stationary (i.e., violating the Markov property), which affects the algorithm convergence~\cite{maddpg}. Evaluation of the single-agent learning approach in a multi-agent environment (denoted by SA-PPO-MA) will be reported in \Cref{subsec:simulation_result}.

Without pursuing a learning system using single-agent DRL any further, a \gls{MA-PPO} algorithm, a multi-agent extension of PPO~\cite{schulman2017proximal}, is developed. MA-PPO fulfills the general principle of centralized training and decentralized execution~\cite{maddpg}, which is essential for reliable MA-DRL. 
In particular, when directly applying a single-agent DRL algorithm in a multi-agent environment (e.g., SA-PPO-MA), each agent simultaneously and independently learns its own value function using local observation, which brings about the non-stationary environment issue. This issue is addressed in MA-PPO by learning a centrally maintained parametric value function $V_{\boldsymbol \omega}$ with the global state input $s_t\in\mathcal{S}$. The value function $V_{\boldsymbol \omega}$ is then shared among all agents. 
% MA-PPO is a multi-agent extension of PPO. Its design follows the same idea as MADDPG~\cite{maddpg}, which extends DDPG~\cite{ddpg} to the multi-agent context. MADDPG needs to learn a centralized Q-function with large function input that contains the multi-agent joint action space. In comparison to MADDPG, MA-PPO is more suitable for training policies since policy training relies only on the V-function with substantially reduced input dimensions. The V-function in MA-PPO is much easier to model as a DNN than the centrally trained Q-function in MADDPG. In view of this, we decide to use MA-PPO to train policies.

% MA-PPO uses a DNN to approximate a value function $V_{\boldsymbol \omega}$ with the global state input $s_t\in \mathcal S$. 
In MA-PPO, $V_{\boldsymbol \omega}$ is approximated by a DNN. Following PPO, $V_{\boldsymbol \omega}$ will be learned in an on-policy fashion by maintaining a collection of network state-transition samples obtained from using the current policies $\{\pi_{\boldsymbol\theta^n}\}_{n=1}^N$. Each state-transition sample $u$ records both global states and agents' local observations:
\begin{equation}
\resizebox{.9\hsize}{!}{$
      u = \left \langle s_t, s_{t+1}, \{\boldsymbol z_t^n\}_{n=1}^N, \{\boldsymbol z_{t+1}^n\}_{n=1}^N, \{\boldsymbol a_t^{n}\}_{n=1}^N, \{r_t^n\}_{n=1}^N \right \rangle
$}
\label{eq:state_transistion_sample}
\end{equation} 
where the global state $s_t$ is for training $V_{\boldsymbol \omega}$ and the local observations $\boldsymbol z_t^n$ is for training $\pi_{\boldsymbol \theta^n}$.
% where $\boldsymbol z_t^n = \{z_t^{n,m}\}_{m=1}^M$, $\boldsymbol z_{t+1}^n = \{z_{t+1}^{n,m}\}_{m=1}^M$, and $\boldsymbol a^{n}_t = \{a_t^{n,m}\}_{m=1}^M$. 

Then several \emph{mini-batches} of samples denoted by $\mathcal{B}$ can be retrieved from the collection to repeatedly train $V_{\boldsymbol \omega}$ to minimize the \emph{Bellman loss} in \eqref{eq:v-function-optim_ma}:
\begin{equation}
        \mathcal{H}(V_{\boldsymbol \omega}) = \frac{1}{\left \| \mathcal{B} \right \|}\sum_{\mathcal{B}} \left ( V_{\boldsymbol \omega}(s_t) - \sum_{n=1}^N r_t^{n} - \gamma V_{\boldsymbol \omega}(s_{t+1}) \right )^2
    \label{eq:v-function-optim_ma}
\end{equation}

Guided by the trained $V_{\boldsymbol \omega}$, each agent in MA-PPO continues to use the sampled mini-batches to update its policy $\pi_{\boldsymbol \theta^n}$ along the direction of the estimated policy gradient $\nabla_{\boldsymbol\theta^n} J(\pi_{\boldsymbol\theta^n})$ in \Cref{subsubsection:policy_cal}.

\subsubsection{Policy Gradient Calculation}\label{subsubsection:policy_cal} 
Following PPO, in order to estimate $\nabla_{\boldsymbol\theta^n} J(\pi_{\boldsymbol\theta^n})$, MA-PPO must find a way to estimate the following gradient:
\begin{equation}
\begin{split}
\nabla_{\boldsymbol\theta^n} \mathcal{L}(\pi_{\boldsymbol\theta^n}) & = \nabla_{\boldsymbol\theta^n} \mathop{\mathbb{E}}_t \left[\frac{\pi_{\boldsymbol \theta^n}(\boldsymbol{a}_t^{n}| \boldsymbol z^n_t)}{\pi_{\boldsymbol \theta^n_{old}}(\boldsymbol{a}_t^{n}| \boldsymbol z^n_t)}A_t(s_t, \{\boldsymbol{a}_t^{n}\}_{n=1}^N) \right]\\
& \approx \frac{1}{\|\mathcal{B}\|} \sum_{\mathcal{B}} \frac{A_t(s_t, \{\boldsymbol{a}_t^{n}\}_{n=1}^N)}{\pi_{\boldsymbol \theta_{old}^n}(\boldsymbol{a}_t^{n}| \boldsymbol z_t^n)} \nabla_{\boldsymbol\theta^n}{ \pi_{\boldsymbol \theta^n}(\boldsymbol{a}_t^n| \boldsymbol z_t^n)}
\end{split}
\end{equation}      
where ${\boldsymbol\theta^n}$ and ${\boldsymbol\theta_{old}^n}$ refer to the policy parameters after and before policy update in a \gls{TI} respectively. $A_t(s_t, \{\boldsymbol{a}_t^{n}\}_{n=1}^N)$ is the advantage function obtained through $V_{\boldsymbol \omega}$ by using the \gls{GAE} technique developed in~\cite{schulman2015high}. 

According to \eqref{eqt:pi_design}, 
\begin{equation*}
a_t^{n,m} - \text{Softmax}(f_{\boldsymbol{\theta}^n}(z^{n,m}_t)) = \epsilon_t^{n,m} \sim \mathcal{N}\left( 0, \sigma^2\right)
\end{equation*}
Therefore, each element $a_t^{n,m}$ in $\boldsymbol a_t^n$ follows a Gaussian distribution:
\begin{equation*}
a_t^{n,m} \sim \mathcal{N}\left( \text{Softmax}(f_{\boldsymbol{\theta}^n}(z^{n,m}_t)), \sigma^2\right)
\end{equation*}
Note that the Gaussian noise $\epsilon_t^{n,m}$ for each $a_t^{n,m}$ is independently sampled. Therefore, 
      \begin{equation*}
      \pi_{\boldsymbol \theta^n}(\boldsymbol{a}^n_t| \boldsymbol z^n_t) = \prod_{m=1}^M \frac{1}{\sigma\sqrt{2\pi}} e^{-\frac{1}{2}(\frac{a_t^{n,m}-\mu_t^{n,m}}{\sigma})^2}
      \end{equation*}
      where $\mu_t^{n,m} = \text{Softmax}(f_{\boldsymbol{\theta}^n}(z^{n,m}_t))$.
       % and $z_t^{n,m} = \mathcal{O}^n(s_t,m)$ is the local observation with respect to each controller $C_{m}$ extracted from $s_t$.

      For each sample $u_t\in \mathcal{B}$, $\nabla_{\boldsymbol\theta^n}{ \pi_{\boldsymbol \theta^n}(\boldsymbol{a}_t^n| \boldsymbol z^n_t)}$ can be calculated by using $\boldsymbol{a}_t^n$ and $\boldsymbol z^n_t$ recorded in sample $u_t$ as shown below:
      \begin{equation}
      \label{eqt:cha5_sadrl_dpi_dtheta}
      \begin{split}
      % \resizebox{.95\hsize}{!}{$
        & \quad \nabla_{\boldsymbol{\theta}^n}{\pi_{\boldsymbol \theta^n}(\boldsymbol{a}_t^n| \boldsymbol z^n_t)}\\
        & = \pi_{\boldsymbol \theta^n}(\boldsymbol{a}_t^n| \boldsymbol z^n_t) \nabla_{\boldsymbol\theta^n}{\log \pi_{\boldsymbol \theta^n}(\boldsymbol{a}_t^n| \boldsymbol z^n_t)}\\
        & = \pi_{\boldsymbol \theta^n}(\boldsymbol{a}_t^n| \boldsymbol z^n_t) \nabla_{\boldsymbol\theta^n}\log \Big( \prod_{m=1}^M \frac{1}{\sigma\sqrt{2\pi}} e^{-\frac{1}{2}(\frac{a_t^{n,m}-\mu_t^{n,m}}{\sigma})^2}\Big) \\
        & = \pi_{\boldsymbol \theta^n}(\boldsymbol{a}_t^n| \boldsymbol z^n_t) \nabla_{\boldsymbol\theta^n} \left( \sum_{m=1}^M \Big(\log \big( e^{-\frac{1}{2}(\frac{a_t^{n,m}-\mu_t^{n,m}}{\sigma})^2}\big)\Big) \right)\\
        & = \pi_{\boldsymbol \theta^n}(\boldsymbol{a}_t^n| \boldsymbol z^n_t) \left( \sum_{m=1}^M \nabla_{\boldsymbol\theta^n} \Big(-\frac{1}{2}(\frac{a_t^{n,m}-\mu_t^{n,m}}{\sigma})^2 \Big)\right)\\
        & = -\frac{\pi_{\boldsymbol \theta^n}(\boldsymbol{a}_t^n| \boldsymbol z^n_t)}{2\sigma^2} \Big(\sum_{m=1}^M \nabla_{\boldsymbol\theta^n} (a_t^{n,m}-\mu_t^{n,m})^2 \Big)\\
        % & = -\frac{\pi_{\boldsymbol \theta}(\boldsymbol{a}_t| \boldsymbol z^n_t)}{2\sigma^2} \Big( \sum_{v'\in V} \nabla_{\boldsymbol\theta} (a_t^{v'}-\mu_t^{n,m})^2 \Big)\\
        & = \frac{\pi_{\boldsymbol \theta^n}(\boldsymbol{a}_t^n| \boldsymbol z^n_t)}{\sigma^2}  \Big( \sum_{m=1}^M (a_t^{n,m}-\mu_t^{n,m}) \nabla_{\boldsymbol\theta^n} \mu_t^{n,m}\Big)
        % \frac{e^{-\frac{1}{2}(\boldsymbol{a}_t-\boldsymbol\mu_t)^T {\boldsymbol \Sigma}^{-1}(\boldsymbol{a}_t-\boldsymbol\mu_t)} }
        % {(2\pi)^{\frac{N}{2}} \operatorname{det}(\boldsymbol \Sigma)^{\frac{1}{2}}} 
      % $}
      \end{split}
      \end{equation}
Given
      \begin{equation*}
      \mu^{n,m}_t = \text{Softmax}(f_{\boldsymbol{\theta}^n}(z^{n,m}_t))
      = \frac{e^{f_{\boldsymbol{\theta}^n}(z^{n,m}_t)}}{\sum_{i=1}^M e^{f_{\boldsymbol{\theta}^n}(z^{n,i}_t)}},
      \end{equation*}
      we have
      \begin{equation}
      \label{eqt:cha5_sadrl_dmu_dtheta}
      % \resizebox{.99\hsize}{!}{$
      \begin{split}
      \nabla_{\boldsymbol\theta^n} \mu_t^{n,m} 
      = &\frac{e^{f_{\boldsymbol{\theta}^n}(z^{n,m}_t)}}
      {(\sum_{i=1}^Me^{f_{\boldsymbol{\theta}^n}(z^{n,i}_t)})^2}
      \sum_{i=1}^M e^{f_{\boldsymbol{\theta}^n}(z^{n,i}_t)} \cdot \\
      & \Big( \nabla_{\boldsymbol{\theta}^n}f_{\boldsymbol{\theta}^n}(z^{n,m}_t)-
      \nabla_{\boldsymbol{\theta}^n}f_{\boldsymbol{\theta}^n}(z^{n,i}_t) \Big)
      \end{split}
      % $}
      \end{equation}
  where $\nabla_{\boldsymbol{\theta}^n}f_{\boldsymbol{\theta}^n}(z^{n,m}_t)$ is the gradient of the priority function (i.e., the DNN) in \Cref{fig:cha5_sa_policy}.

  Summarizing the above discussions, with respect to a mini-batch $\mathcal{B}$, $\nabla_{\boldsymbol{\theta}^n} \mathcal{L}(\pi_{\boldsymbol{\theta}^n})$ is estimated using \eqref{eqt:cha5_sappo_policy_gradient}:
      \begin{equation}\label{eqt:cha5_sappo_policy_gradient}
      \resizebox{.99\hsize}{!}{$
      \begin{split}
       &\nabla_{\boldsymbol{\theta}^n} \mathcal{L}(\pi_{\boldsymbol{\theta}^n})
       \approx \frac{1}{\left \| \mathcal{B} \right \|} \sum_{\mathcal{B}}
      \frac{A_t(s_t, \{\boldsymbol{a}^n_t\}_{n=1}^N)}
      {\pi_{\boldsymbol \theta_{old}^n}(\boldsymbol{a}^n_t| \boldsymbol z^n_t)} \cdot
      \frac{\pi_{\boldsymbol \theta^n}(\boldsymbol{a}^n_t| \boldsymbol z^n_t)}{\sigma^2}  \left( \sum_{m=1}^M (a_t^{n,m}-\mu_t^{n,m}) \cdot \right. \\
        & \left. \frac{e^{f_{\boldsymbol{\theta}^n}(z^{n,m}_t)}}
      {(\sum_{i=1}^Me^{f_{\boldsymbol{\theta}^n}(z^{n,i}_t)})^2}
      \sum_{i=1}^Me^{f_{\boldsymbol{\theta}^n}(z^{n,i}_t)}
      \Big( \nabla_{\boldsymbol{\theta}^n}f_{\boldsymbol{\theta}^n}(z^{n,m}_t)-
      \nabla_{\boldsymbol{\theta}^n}f_{\boldsymbol{\theta}^n}(z^{n,i}_t) \Big)\right)
        \end{split}
        $}
      \end{equation}
provided that $\frac{\pi_{\boldsymbol\theta^n}}{\pi_{\boldsymbol\theta_{old}^n}}$ falls in the range $(-\infty,1+\varepsilon\footnote{$\varepsilon$ is a hyper-parameter that is set to $0.2$ following PPO.})$ if $A_t(s_t, \{\boldsymbol{a}^n_t\}_{n=1}^N))>0$ or $(1-\varepsilon,+\infty)$ if $A_t(s_t, \{\boldsymbol{a}^n_t\}_{n=1}^N))<0$, with respect to any $\{\boldsymbol{a}^n_t\}_{n=1}^N$ and $s_t$. Otherwise,~$\nabla_{\boldsymbol\theta^n} \mathcal{L}(\pi_{\boldsymbol\theta^n}) = 0$. 

According to PPO, the policy $\pi_{\boldsymbol \theta^n}$ can be improved by repeatedly updating the policy parameters $\boldsymbol \theta^n$ along the direction of $\nabla_{\boldsymbol{\theta}^n} \mathcal{L}(\pi_{\boldsymbol{\theta}^n})$. Note that this technique for calculating the policy gradient can be easily extended to the case with an arbitrary number of controllers. With the help of TensorFlow, the gradient calculation can also be fully automated in our training system, regardless of how many controllers are involved. The computational complexity is linear with respect to the number of controllers. 
% The overall training process is summarized in Appendix~\ref{alg:mappo_training}. 
  
\section{Simulation} \label{sec:simulation}
  In this section, we first introduce the simulation setting which includes the algorithm implementation and the network simulation setting. To demonstrate the effectiveness of the new policy design, evaluating it under a single-agent DRL framework is more preferable compared to a multi-agent DRL framework. This is mainly because the performance of MA-DRL depends not only on the policy design but also other factors, such as inter-agent cooperation and non-stationary environment handling. In view of this, it is easier and more straightforward to demonstrate the effectiveness of the new policy design in a single-agent DRL setting where it is compared with a non-adaptive policy\footnote{The non-adaptive policy uses the traditional policy design where the policy is directly represented as a DNN.}. 
  % Detailed DNN settings for the two policies are provided in Appendix~\ref{sec:nn_architectureSA}.
  Similar to~\cite{jay2019deep}, we also investigate the influence of historical information and the discount factor $\gamma$ on the performance respectively. 

  After that, simulations are conducted to demonstrate the necessity of using MA-DRL for policy training in a multi-agent environment. In particular, the policy trained by MA-PPO is compared with the policy trained by each agent independently using single-agent training approach (denoted by SA-PPO-MA) as we discussed in \Cref{subsec:ma_ppo}. 

  To demonstrate the effectiveness of the policy trained by MA-PPO, it is also compared with a widely used man-made policy weighted round robin (denoted by CWRR), a recently proposed GD-based policy (denoted by GD)~\cite{huang2020scalable}, and the centralized single-agent policy (denoted by Central) with full observability of the environment. 
  % All the algorithms that are evaluated in this section are summarized in Appendix~\ref{sec:comparedAlg}.
  In terms of comparisons with other MA-DRL algorithms, MA-DDPG is closely related to MA-PPO. However, deterministic policy gradient used in MA-DDPG cannot be calculated with respect to our adaptive policy network design, rendering MA-DDPG inapplicable. Apart from that, the new technique developed in \Cref{subsubsection:policy_cal} to compute the gradient of our policy network can be utilized by any AC algorithms designed for training stochastic policies such as TRPO~\cite{trpo} and Asynchronous Advantage Actor-Critic (A3C)~\cite{mnih2016asynchronous}. 
  However, investigating the performance of different AC algorithms is not the main focus of this paper. Furthermore, compared to PPO, TRPO has high computation complexity due to its use of both linear approximation of the learning objective and quadratic approximation of the constraint for policy update~\cite{trpo,chen2018adaptive}. On the other hand, A3C asynchronously executes multiple actors where each actor interacts with its own copy of the environment. The use of multiple actors inevitably requires more computation resources. 
  % The rewards collected from all actors are used to update the shared policy and value function. However, due to the use of multiple actors, A3C requires more computation resources. 
  Apart from that, the policy updates in A3C rely on the latest data collected from multiple actors without using memory replay, which results in high sampling costs. Therefore, both TRPO and A3C are not as suitable as PPO. Moreover, as a representative algorithm among all AC algorithms, studying the performance of PPO gives us an overall good understanding of other AC algorithms. In the future, combining our new policy design with different AC algorithms in a multi-agent setting will be investigated when enough computation resources and time are provided.

    % Specifically, the non-adaptive policy uses the same NN configuration as our new policy design except for the number of input and output nodes. The number of output nodes for the non-adaptive policy is the number of available controllers and the number of input nodes depends on the state $s_t$.  

    \subsection{Algorithm Implementation} \label{subsec:algorithm_setting}
    We implement MA-PPO based on the high-quality implementation of PPO provided by OpenAI baselines\footnote{https://github.com/openai/baselines}. 
    To identify the suitable NN architecture for both the priority function $f_{\boldsymbol{\theta}^n}$ and value function $V_{\boldsymbol{\omega}}$, different NN architectures are compared to see their impacts on the network performance. Based on our preliminary study, a fully connected multilayer feed forward NN with two hidden layers of 64 ReLU units is adopted for both $f_{\boldsymbol{\theta}^n}$ and $V_{\boldsymbol{\omega}}$, which is also the same NN architecture recommended in PPO. 
    % In our simulation, we adopt a common NN architecture recommended in PPO for both the priority function $f_{\boldsymbol{\theta}^n}$ and value function $\mathcal{V}_{\boldsymbol{\omega}}$. In particular, each NN is a fully connected multilayer feed forward neural network with two hidden layers of 64 ReLU units and linear activation is used for the output layer. More discussions about selecting the NN architecture can be found in Appendix~\ref{sec:nn_architecture}. 

    Meanwhile, we follow closely the hyper-parameter settings of PPO on Mujoco benchmarks in~\cite{schulman2017proximal}. However, there are a few exceptions. Specifically, the Gaussian noises $\boldsymbol{\epsilon}_t$ in~\eqref{eqt:rho_sa} have their standard deviations set to 0.01. During every algorithm run, the policy is trained for $900$ \gls{TI}s which consist of 1800 episodes and each episode contains 60 time steps. 
    % In each \gls{TI}, two episodes are simulated with two different request arrival rates representing the low workload and high workload scenarios respectively to enable the agent to learn how to dispatch requests under different workloads (see \Cref{cha5_subsubsec:network_simulation_setting} for the workload setting). Given a fixed arrival rate, the network simulates one episode and each episode contains $60$ time steps which is $30$ minutes of the corresponding simulated network operation time (see \Cref{cha5_subsubsec:network_simulation_setting} for more discussions).
    % each iteration contains 2 network settings with same topology but different request arrival rates. Each network setting simulates one episode and each episode contains $60$ time steps. 
    Both $\boldsymbol{\theta}^n$ and $\boldsymbol{\omega}$ are trained using data sampled from the current \gls{TI}. The NN parameters $\boldsymbol{\theta}^n$ and $\boldsymbol{\omega}$ are updated using Adam optimizer with $3 \times 10^{-4}$ learning rate, $40$ minibatch size, and~$8$ epochs. 

    \subsection{Network Simulation Setting}\label{subsec:network_simulation_setting}
    Simulations are conducted under real network topologies provided by Sprint~\cite{sprint}: South America and Asia Sprint networks equipped with 8 and 14 switch centers respectively.
    % , as shown in 
    % % \Cref{fig:chap5_network_topo}.
    % Appendix~\ref{sec:network_topology}. 
    A set of heterogeneous controllers with capacities ranging from 6000 pkt/s to 9000 pkt/s have been deployed into the network using existing controller placement algorithm~\cite{huang2020scalable}. Unless we explicitly specify, the numbers of controllers deployed in the South America and Asia networks are 3 and 4 respectively during the simulation. For the centralized single-agent policy (i.e., Central), the location of the centralized agent is selected so that the average propagation latency between the agent and all controllers is minimized. For SA-PPO-MA and MA-PPO, a separate agent is placed for each switch center in the network. 
    % At the beginning of any time step, all agents run their policies individually to calculate their respective priorities of dispatching any new requests to each controller in the network for the next time step.

    % \begin{figure}[]
    %   \begin{center}
    %     \subfloat[South America]{\label{fig:chap5_southAm_topo}\includegraphics[width=.45\linewidth]{figures/chapter5/Sprint_southAmerica.pdf}} \\
    %     \subfloat[Asia]{\label{fig:chap5_asia_topo}\includegraphics[width=.7\linewidth]{figures/chapter5/Sprint_asia.pdf}}
    %   \end{center} 
    % \caption{Network topologies used in simulation studies, obtained from Sprint~\cite{sprint}.}
    %     \label{fig:chap5_network_topo}
    %   \end{figure}

    Each episode is initialized with $0\%$ utilization for all controllers and $0$ packets in the network. During our simulation, the requests arriving at each agent follow the Poisson distribution. CWRR is used to make RD decisions during the warm-up period. The warm-up period lasts for 30 simulated seconds which is assumed to be sufficiently long for the network to enter and stay in a stationary condition. 
    Each simulation episode runs for 30 simulated minutes which is divided into a series of time steps. Every time step lasts for 30 consecutive simulated seconds. At the beginning of each time step, each agent executes its policy individually to calculate the priority of dispatching any new requests to each controller in the network for the next time step, i.e., the next 30 simulated seconds.

    To enable the agent to learn how to dispatch requests under different workloads, two episodes with two request arrival rates are simulated in each \gls{TI}. In particular, for the low workload setting, the combined request arrival rate from all switches is set to be 50$\%$ of the total control plane capacity while the arrival rate under high workload is 80$\%$. 
    % In the simulation result section, the two workloads used for training are marked as ``LWL'' for low workload and ``HWL'' for high workload respectively.

    For the policy to work properly, each agent must provide its local observations $\{z_t^{n,m}\}_{m=1}^M$ to the priority function $f_{\boldsymbol\theta^n}$ in \Cref{fig:cha5_sa_policy}. In consideration of the importance of controllers' capacities, their distance and current availability, as well as the communication demand experienced by the agent, the local observation $z_t^{n,m}$ with respect to $C_{m}$ consists of the following network statistics: (1) request arrival rate history of the switch center $Sw_n$; (2) the processing capacity of $C_{m}$; (3) the propagation latency between $Sw_n$ and~$C_{m}$; (4) the queue length of $C_{m}$; (5) the number of requests sent from $Agt_n$ to $C_{m}$ during the previous time step; (6) the total number of requests received by $C_{m}$ during the previous time step.

    In practice, the request arrival history is made up of a list of request arrival rates measured in the past few time steps by the agent. Intuitively, the longer the list, the easier it is for the agent to detect traffic change patterns and adjusts its request dispatching in consideration of future communication demand. Moreover, an observation with a longer historical list provides more information of the past, which can better fulfill the Markov property. The impact of the history length will be investigated in \Cref{subsubsec:hist_gamma_impact}.

    Similar to $z_t^{n,m}$, the global state $s_t$ contains the arrival rate history from the data plane, all controllers' processing capacity, all controllers' queue length, and the propagation latency measured in $\boldsymbol D$. 

    \subsection{Simulation Result}\label{subsec:simulation_result}

    \subsubsection{Effectiveness of the Adaptive Policy Design} 
\begin{figure}[!tb]
      \begin{center}
        \subfloat[South America]{\label{fig:chap5_southAm_SA_diffalgo}\includegraphics[width=.5\linewidth]{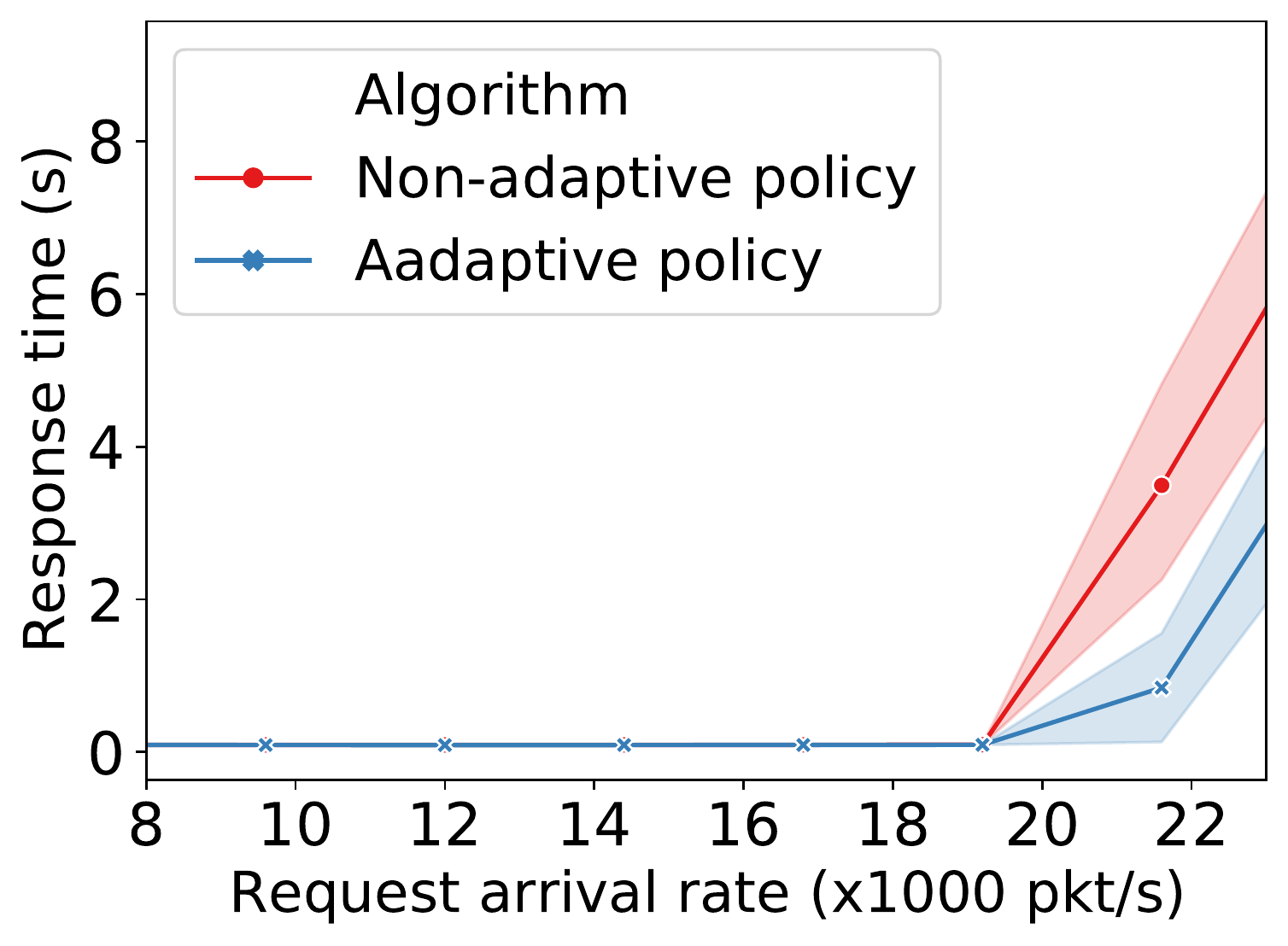}} % \enspace
        \subfloat[Enlarged bottom area of (a)]{\label{fig:chap5_southAm_SA_diffalgo_enlarge}\includegraphics[width=0.5\linewidth]{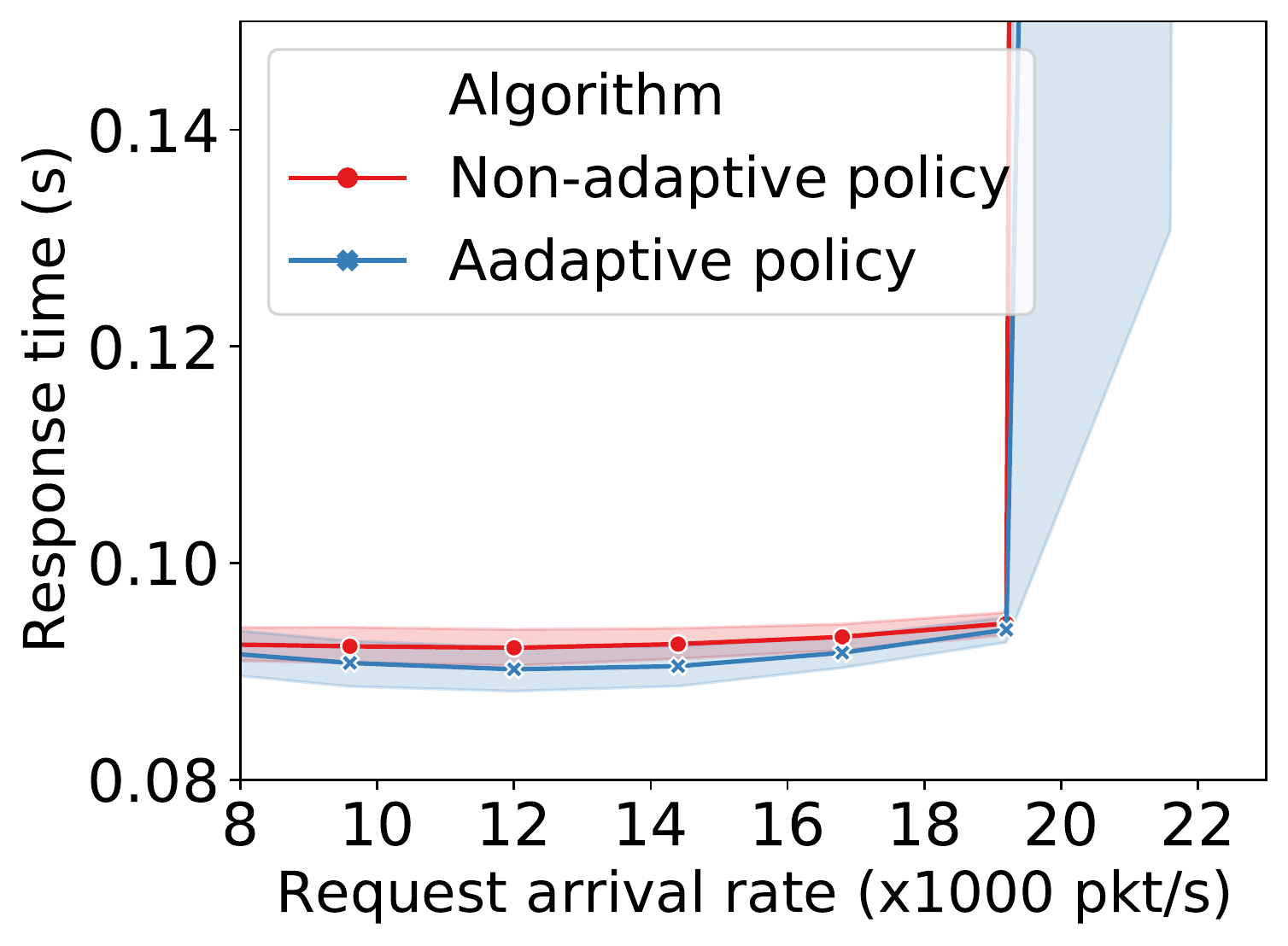}} \\
        \subfloat[Asia]{\label{fig:chap5_asia_SA_diffalgo}\includegraphics[width=.5\linewidth]{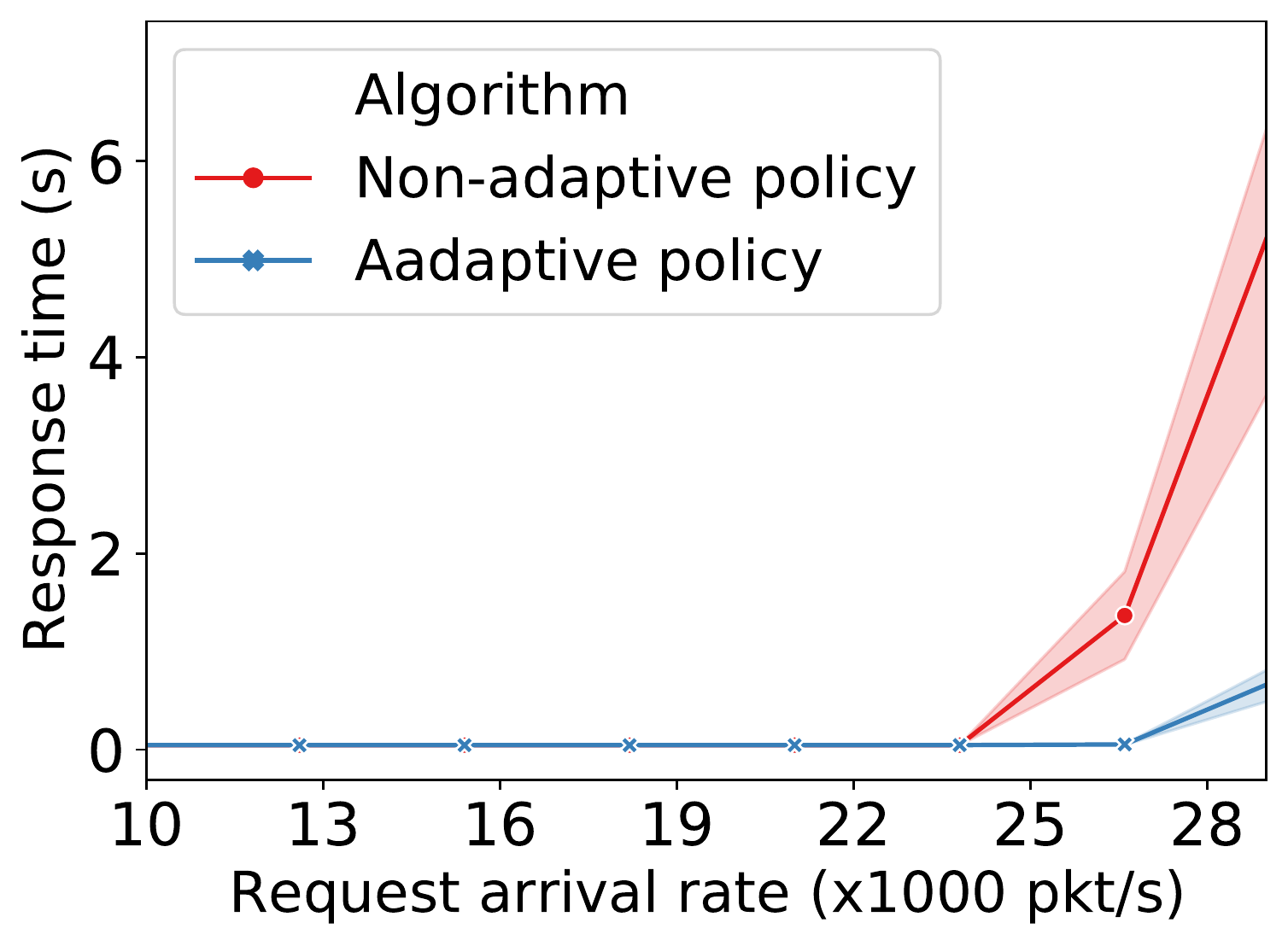}}
        \subfloat[Enlarged bottom area of (c)]{\label{fig:chap5_asia_SA_diffalgo_enlarge}\includegraphics[width=0.5\linewidth]{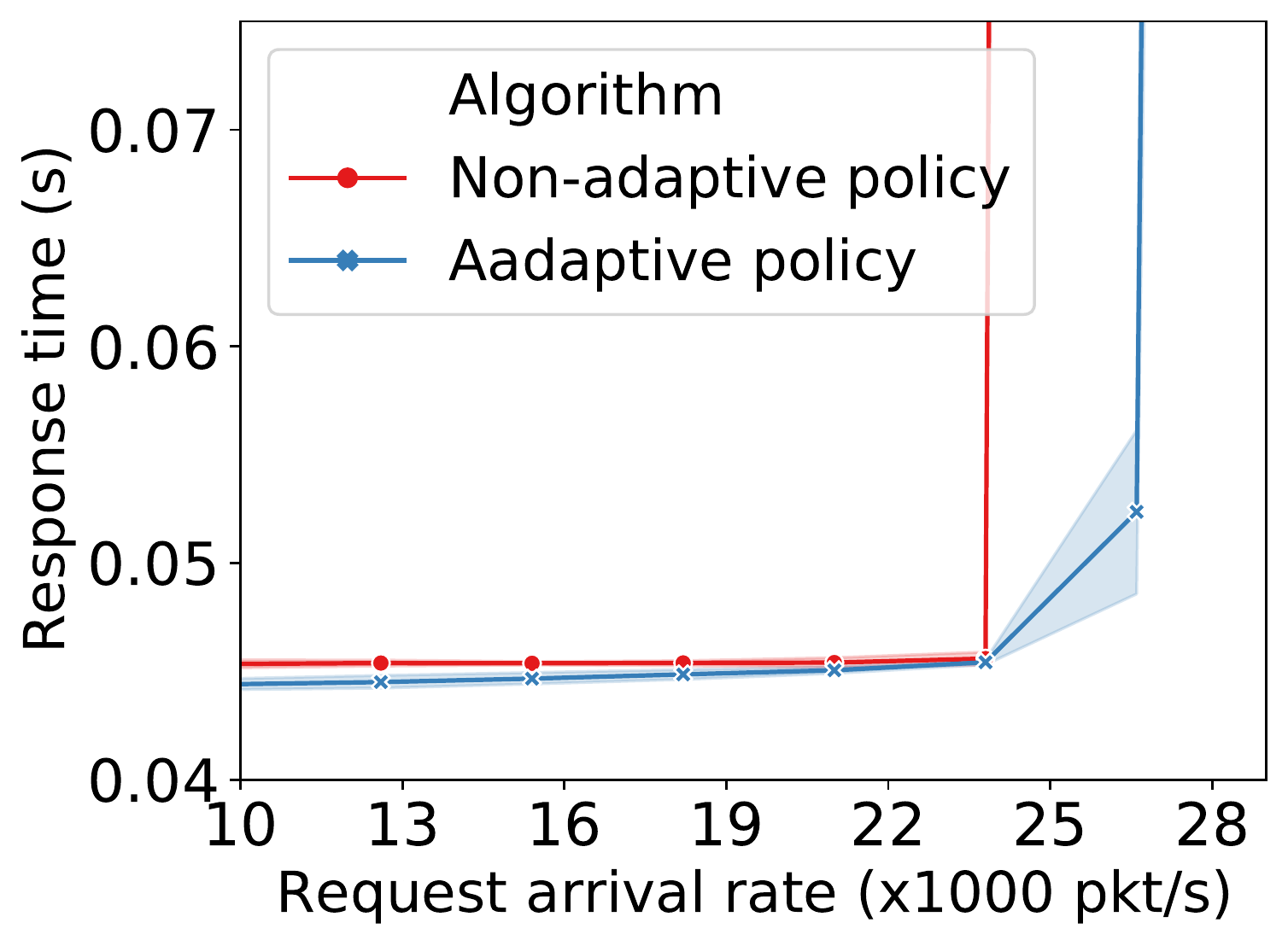}} \\
      \end{center} 
    \caption{Comparison of different policy representations (adaptive vs. non-adaptive) in two network topologies.}
    \label{fig:chap5_SA_perf_diffalgo}
    \end{figure}

    \textbf{Adaptive vs. Non-adaptive policy designs}:
    As shown in Figure~\subref{fig:chap5_southAm_SA_diffalgo_enlarge} and Figure~\subref{fig:chap5_asia_SA_diffalgo_enlarge}, our adaptive policy achieves similar performance as the non-adaptive policy under low request arrival rates. However, from Figure~\subref{fig:chap5_southAm_SA_diffalgo}, we can spot a sudden growth in response time for the non-adaptive policy as the request arrival rate increases while the response time of our policy remains low. This is mainly because in the non-adaptive policy representation, the NN is designed to directly output the request dispatching probabilities over all controllers given all controllers' state information.
    % as listed in Appendix~\ref{sec:nn_architectureSA}. 
    On the other hand, the NN used in our policy is designed to estimate a priority value with respect to one controller using the controller's state information. Given the larger dimensions of both inputs and outputs, the mapping learned in the non-adaptive policy is more complicated than in our policy. Therefore, an NN with the same hidden layer configuration as our policy may not be powerful enough to capture the mapping. This can be further evidenced in Asia topology with more network nodes (Figure~\subref{fig:chap5_asia_SA_diffalgo}) where the performance difference at high request arrival rates is more significant than in the South America topology (Figure~\subref{fig:chap5_southAm_SA_diffalgo}). Therefore, our new policy design can reduce the NN complexity without performance compromise.

    \begin{figure}[!tb]
      \begin{center}
        \subfloat[Testing in a 6-controller network]{\label{fig:chap5_SA_6ctl}\includegraphics[width=.5\linewidth]{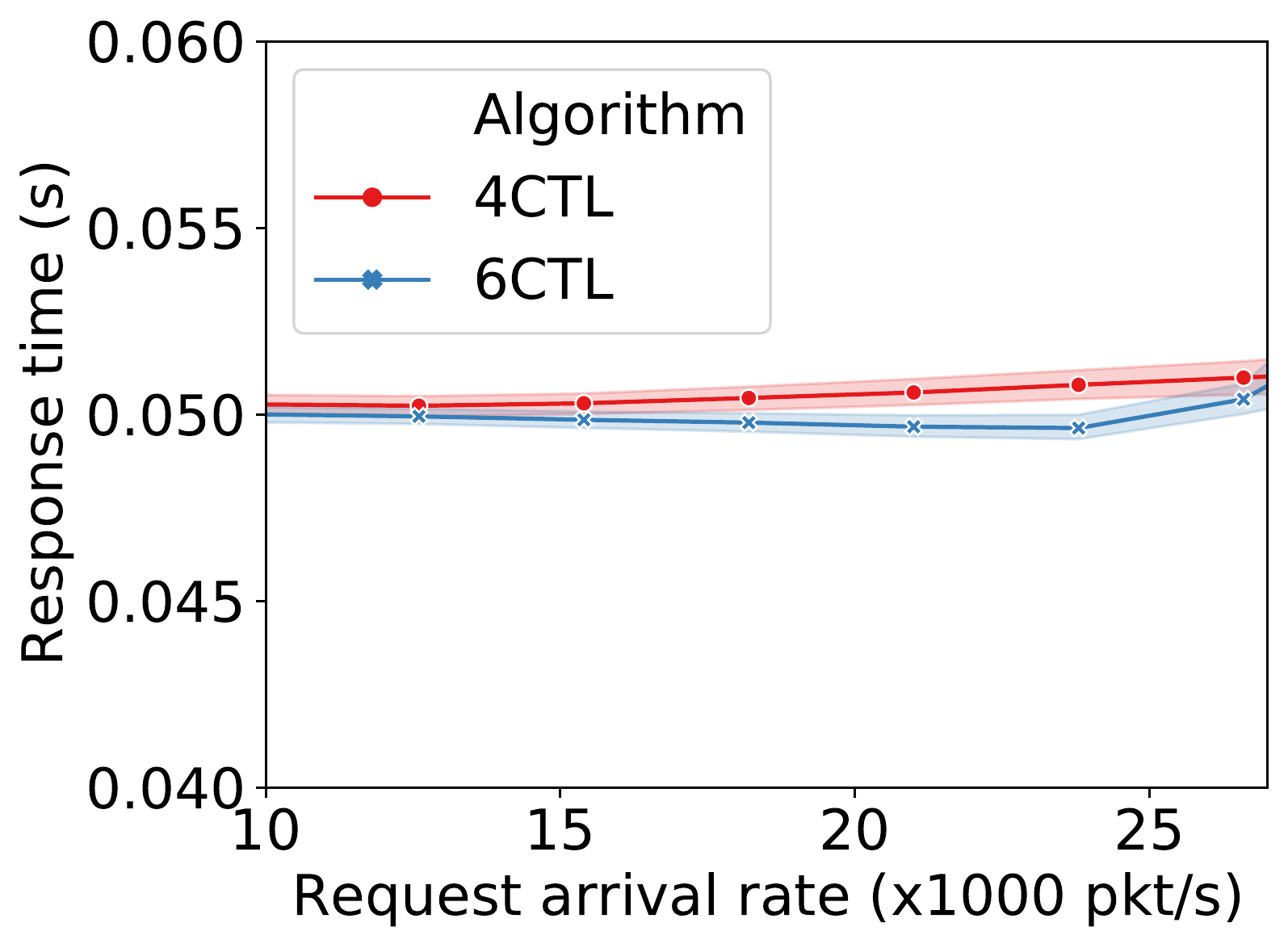}}
        \subfloat[Testing in a 4-controller network]{\label{fig:chap5_SA_4ctl}\includegraphics[width=.5\linewidth]{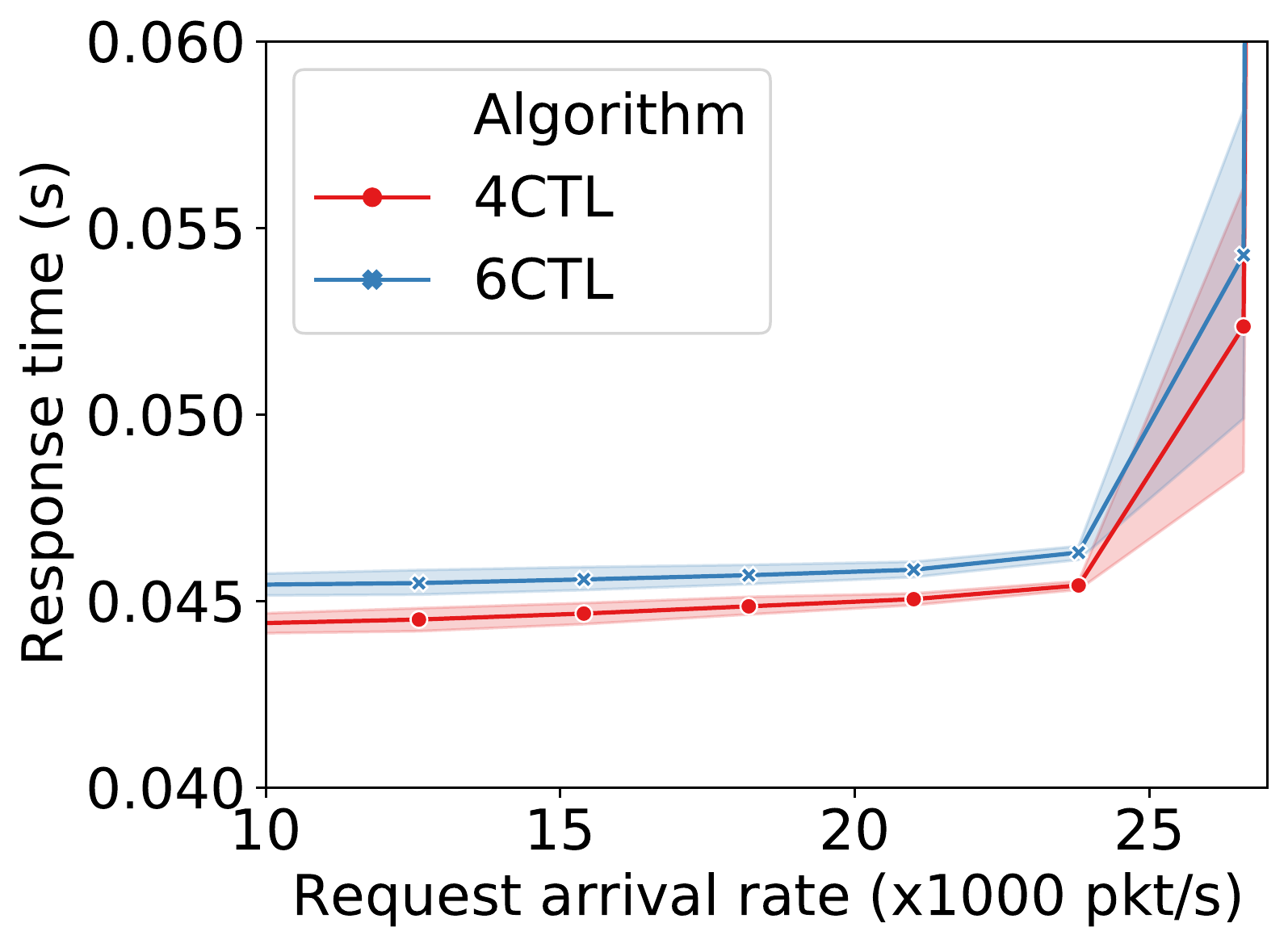}} % \enspace
        \end{center}
    \caption{Policy adaptiveness with a changing number of controllers. $X$CTL refers to the policy trained in a network with $X$ controllers.}
    \label{fig:chap5_SA_perf_diffctl} 
    \end{figure}

    \textbf{Policy adaptiveness}: Although our policy was only trained under two different workloads (50\% and 80\%), it can perform consistently well under different workloads, ranging from 30\% up to 90\% as demonstrated in \Cref{fig:chap5_SA_perf_diffalgo}. 

    To demonstrate the adaptiveness of our trained policy with respect to different numbers of controllers, the policy trained in a network with 4 controllers (4CTL) is evaluated in a network with 6 controllers. Its performance is compared with the policy trained with 6 controllers (6CTL). From Figure~\subref{fig:chap5_SA_6ctl}, we can see that 4CTL can achieve similar performance compared to 6CTL. Similar conclusions can also be drawn from Figure~\subref{fig:chap5_SA_4ctl} where 6CTL is compared with 4CTL in a network with 4 controllers. Our simulation results confirm that the policy can perform consistently well in networks with changing numbers of controllers.

        \begin{figure}[!tbp]
      \begin{center}
        \subfloat[]{\label{fig:chap5_southAm_SAtest_hist}\includegraphics[width=.5\linewidth]{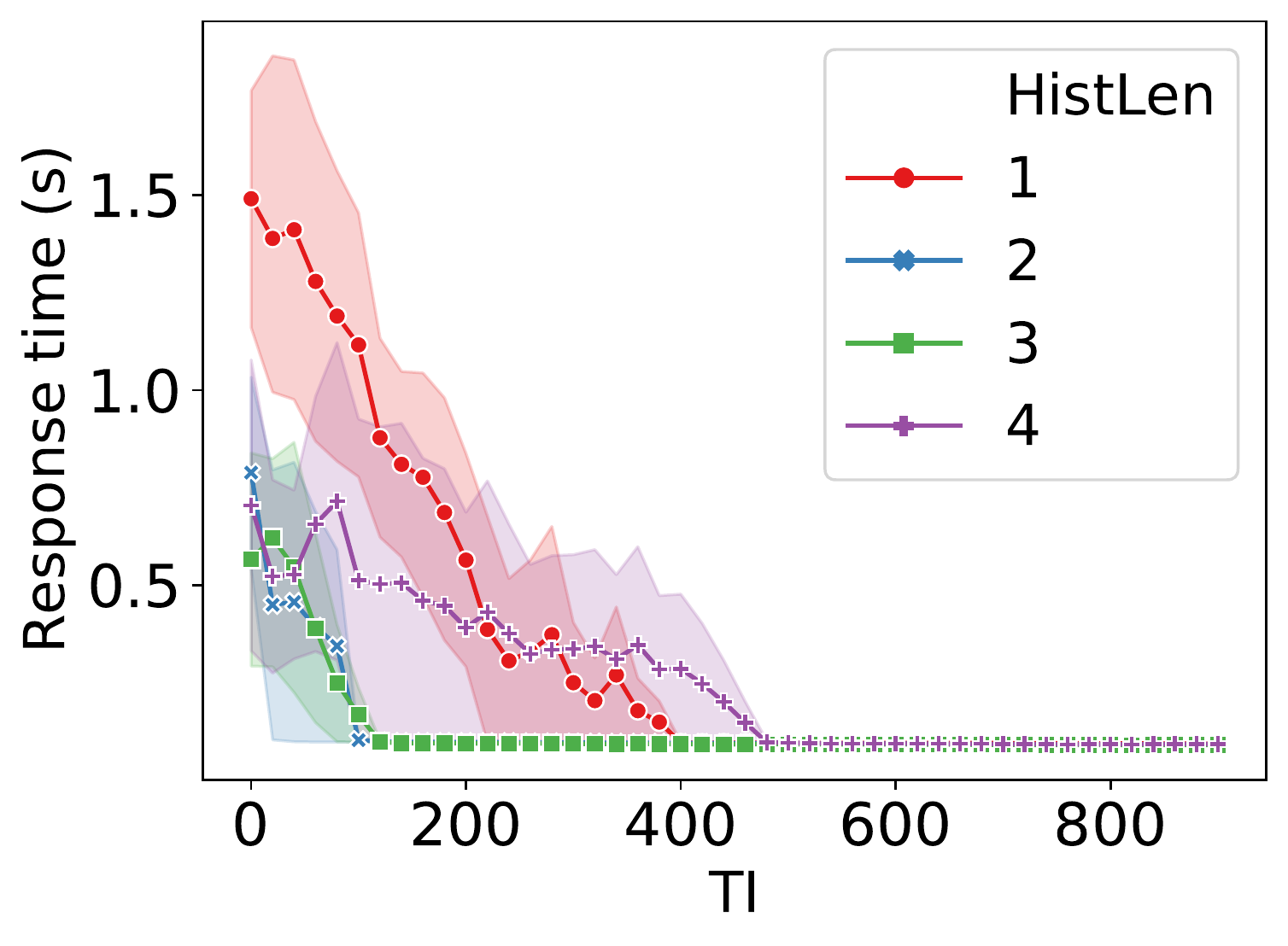}} \enspace
        \subfloat[Enlarged bottom area of (a)]{\label{fig:chap5_southAm_SAtest_hist_enlarge}\includegraphics[width=.5\linewidth]{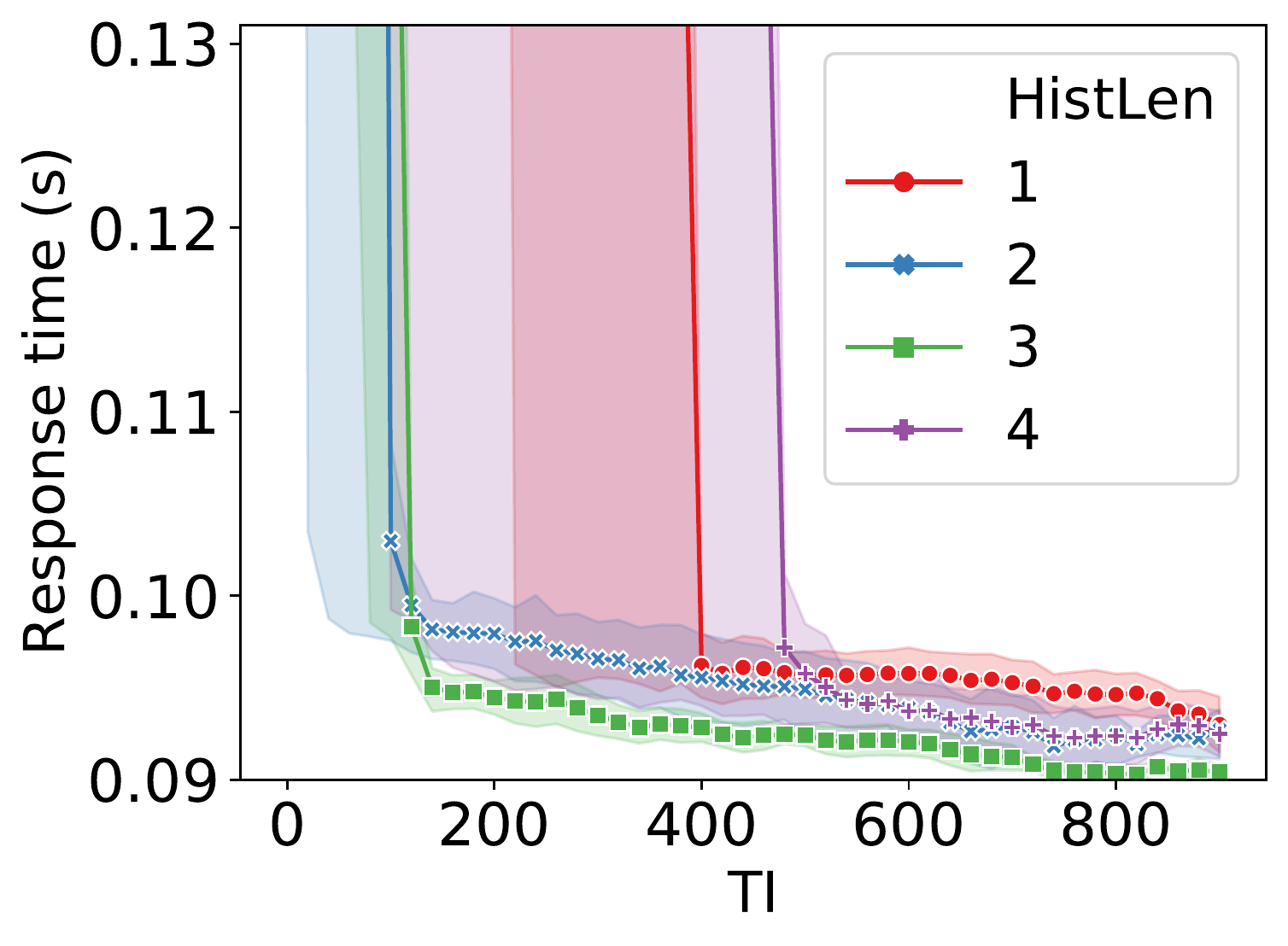}}
      \end{center} 
    \caption{Influence of historical information.
    }
    \label{fig:chap5_SAtest_perf_hist}
    \end{figure}

    \subsubsection{Performance Impact of History Length and $\gamma$}\label{subsubsec:hist_gamma_impact} 
    \textbf{Impact of history length}: Similar to~\cite{jay2019deep}, we investigate the influence of historical information on the performance of our policy. As shown in \Cref{fig:chap5_SAtest_perf_hist}, regarding the list of historical request arrival rates contained in the agent's observation, its length needs to be set properly. With a larger history length, more information of the past is included in the agent's observation, which provides a better approximation of a Markov state. However, if the length is too large (e.g., 4), more learning samples are required for the network to improve its performance. On the other hand, when the length is too small (e.g., 1), the response time stops reducing after 400 TI. It appears that the most suitable length is 3 in our simulations for a good trade-off between sampling costs and performance. 

    \begin{figure}[!tb]
      \begin{center}
        \subfloat[South America]{\label{fig:chap5_southAm_SAtest}\includegraphics[width=.5\linewidth]{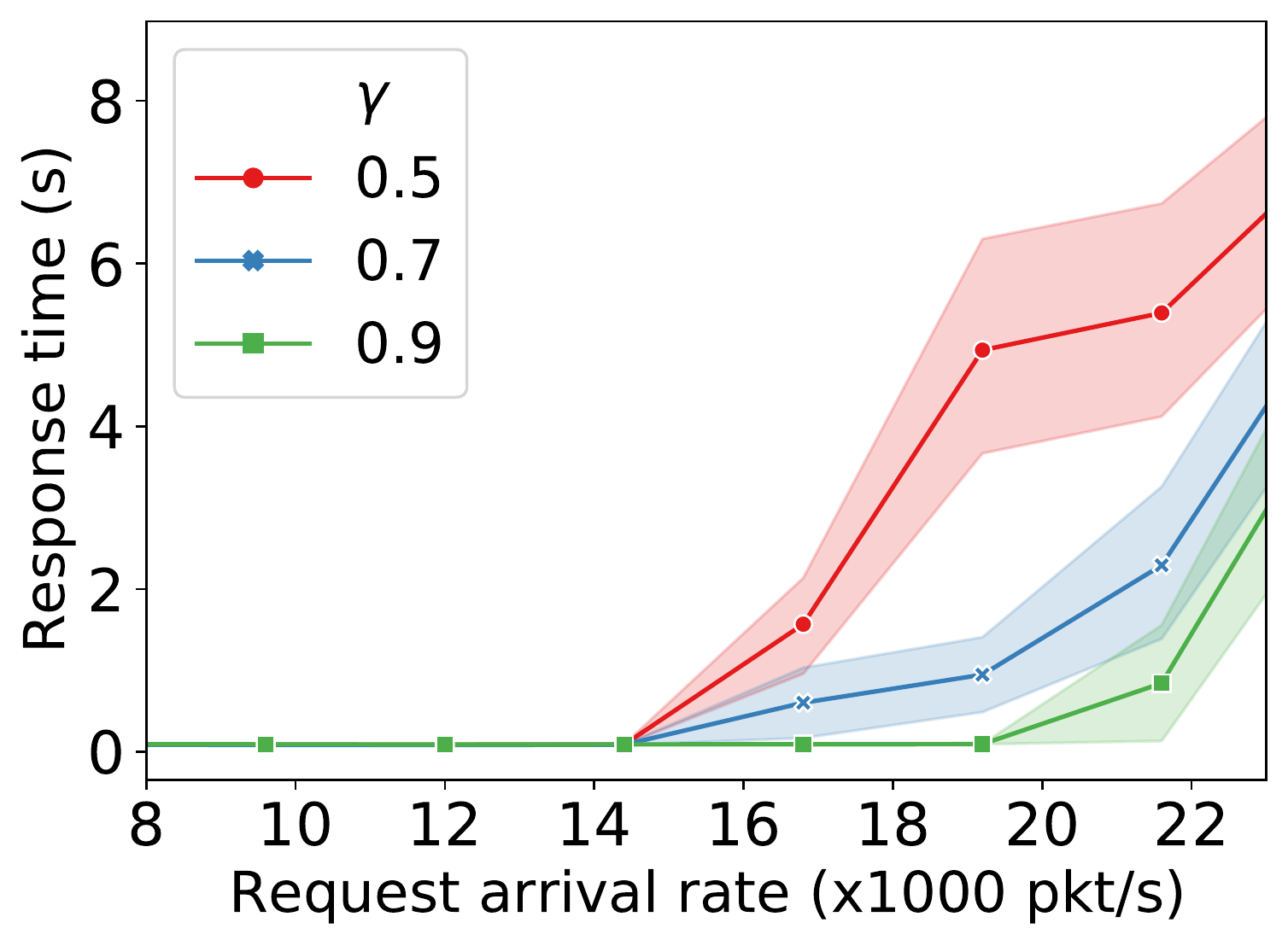}} \enspace
        \subfloat[Asia]{\label{fig:chap5_asia_SAtest}\includegraphics[width=.5\linewidth]{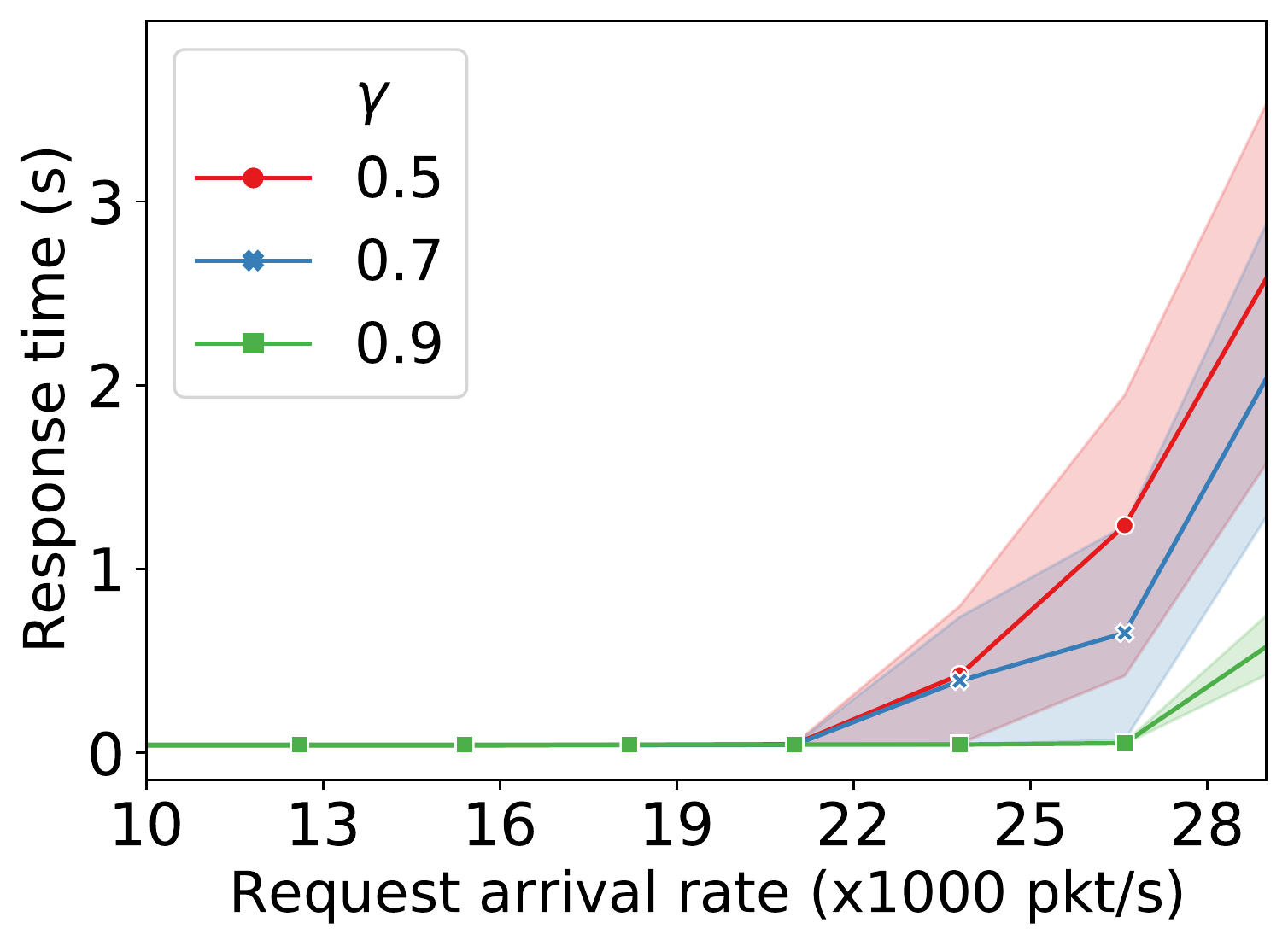}}
      \end{center} 
    \caption{Influence of $\gamma$. 
    % In particular, ``LWL'' and ``HWL'' indicate the two request arrival rates used during the training.
    }
    \label{fig:chap5_SAtest_perf_sadrl_gamma}
    \end{figure}

    \textbf{Impact of $\gamma$}: We also investigate the influence of $\gamma$ on the performance of our policy. 
    \Cref{fig:chap5_SAtest_perf_sadrl_gamma} demonstrates the evaluation of the trained policies with different $\gamma$ under a broad range of request arrival rates in two topologies. From \Cref{fig:chap5_SAtest_perf_sadrl_gamma}, we can see that the policy with $\gamma=0.9$ consistently achieves the lowest response time compared to policies with $\gamma=0.5$ and $\gamma=0.7$ in both topologies. It confirms our theory that the agent should consider the influence of its actions on future network performance, which is vital to prevent any controllers from being overloaded due to accumulated requests over a long run. Thus, for the remaining simulation studies, $\gamma$ is fixed to 0.9. Apart from that, we also observe that as the request arrival rate exceeds a certain value, the response time of all policies increases sharply regardless of the $\gamma$ values. This is mainly because the control plane is highly loaded.

  \subsubsection{Effectiveness of MA-PPO}\label{subsubsec:ma_simulation} 

    \begin{figure}[!tb]
      \begin{center}
        \subfloat[South America]{\label{fig:chap5_southAm_MAtest_gamma09}\includegraphics[width=0.46\linewidth]{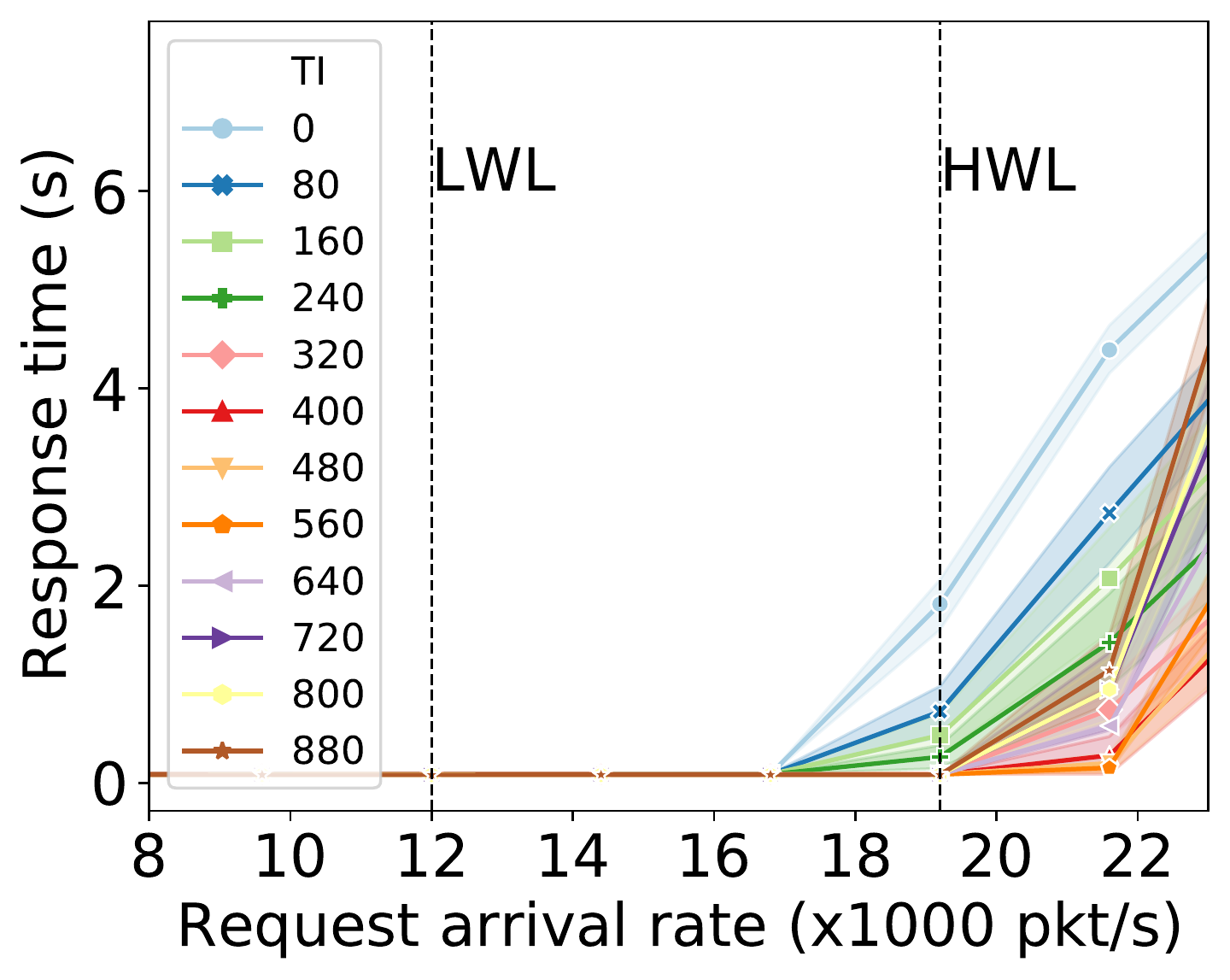}} %\enspace
        \subfloat[Enlarged bottom area of (a)]{\label{fig:chap5_southAm_MAtest_gamma09_enlarge}\includegraphics[width=0.5\linewidth]{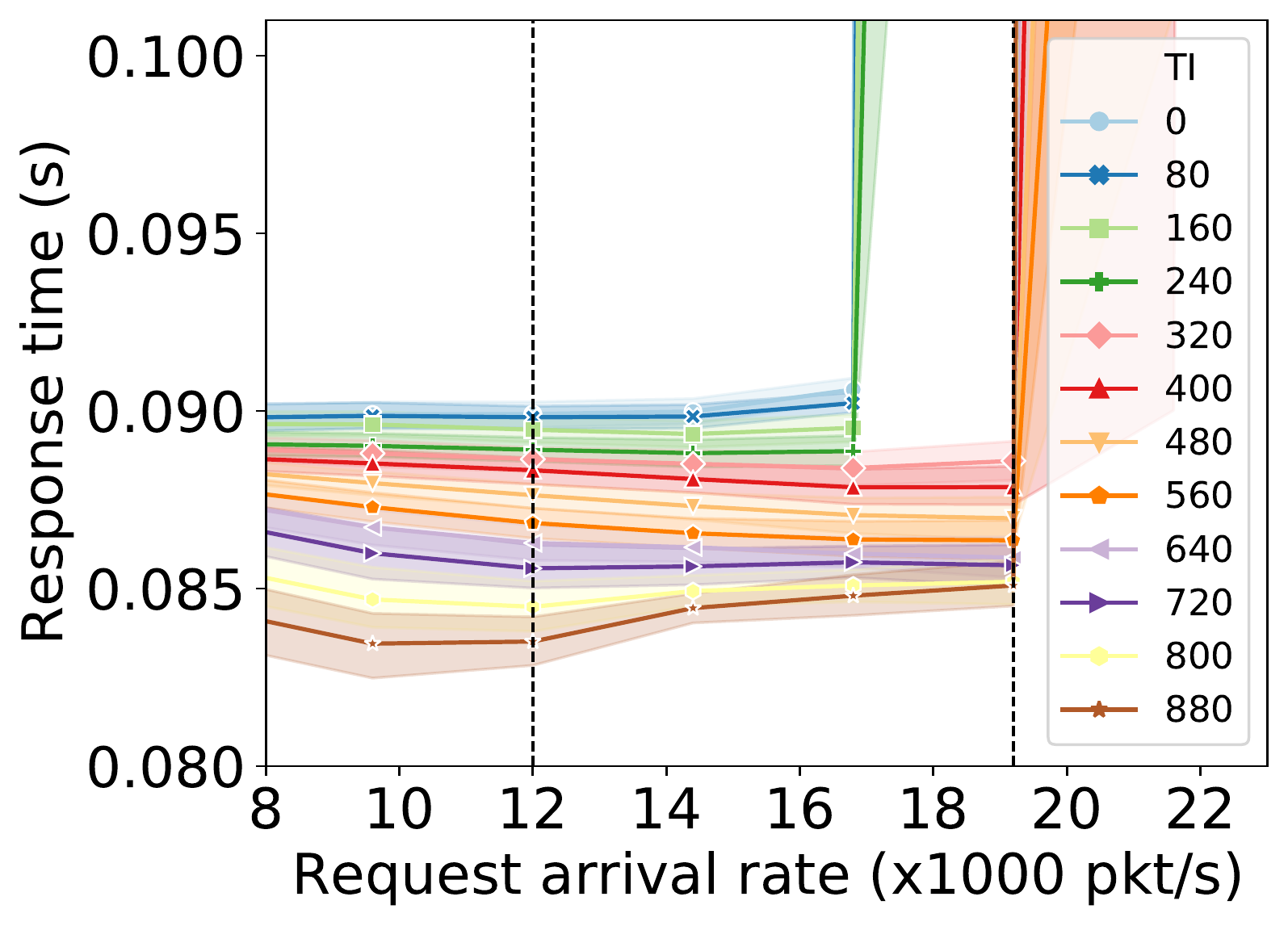}}\\
        \subfloat[Asia]{\label{fig:chap5_asia_MAtest_gamma09}\includegraphics[width=0.46\linewidth]{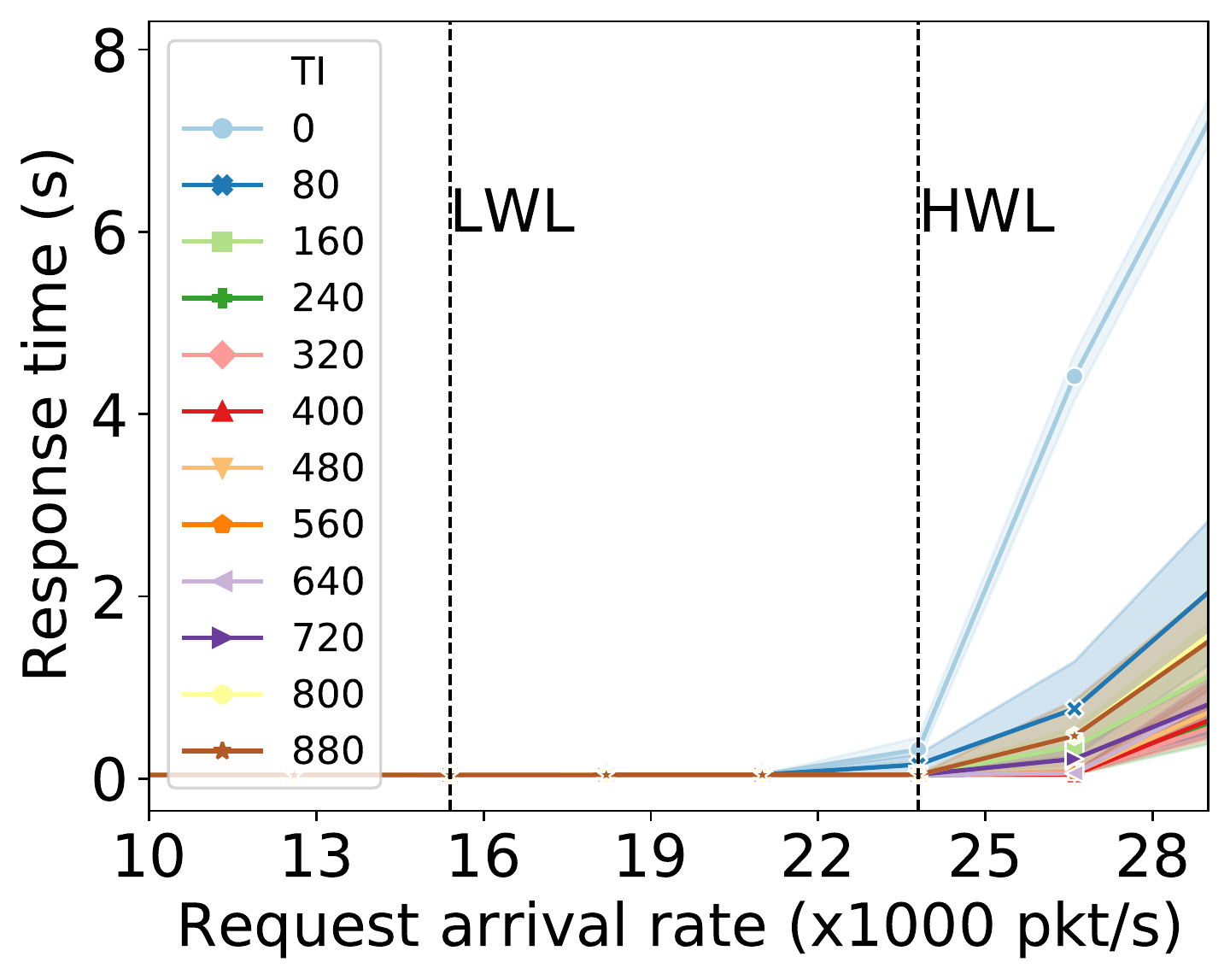}} %\enspace
        \subfloat[Enlarged bottom area of (c)]{\label{fig:chap5_asia_MAtest_gamma09_enlarge}\includegraphics[width=0.5\linewidth]{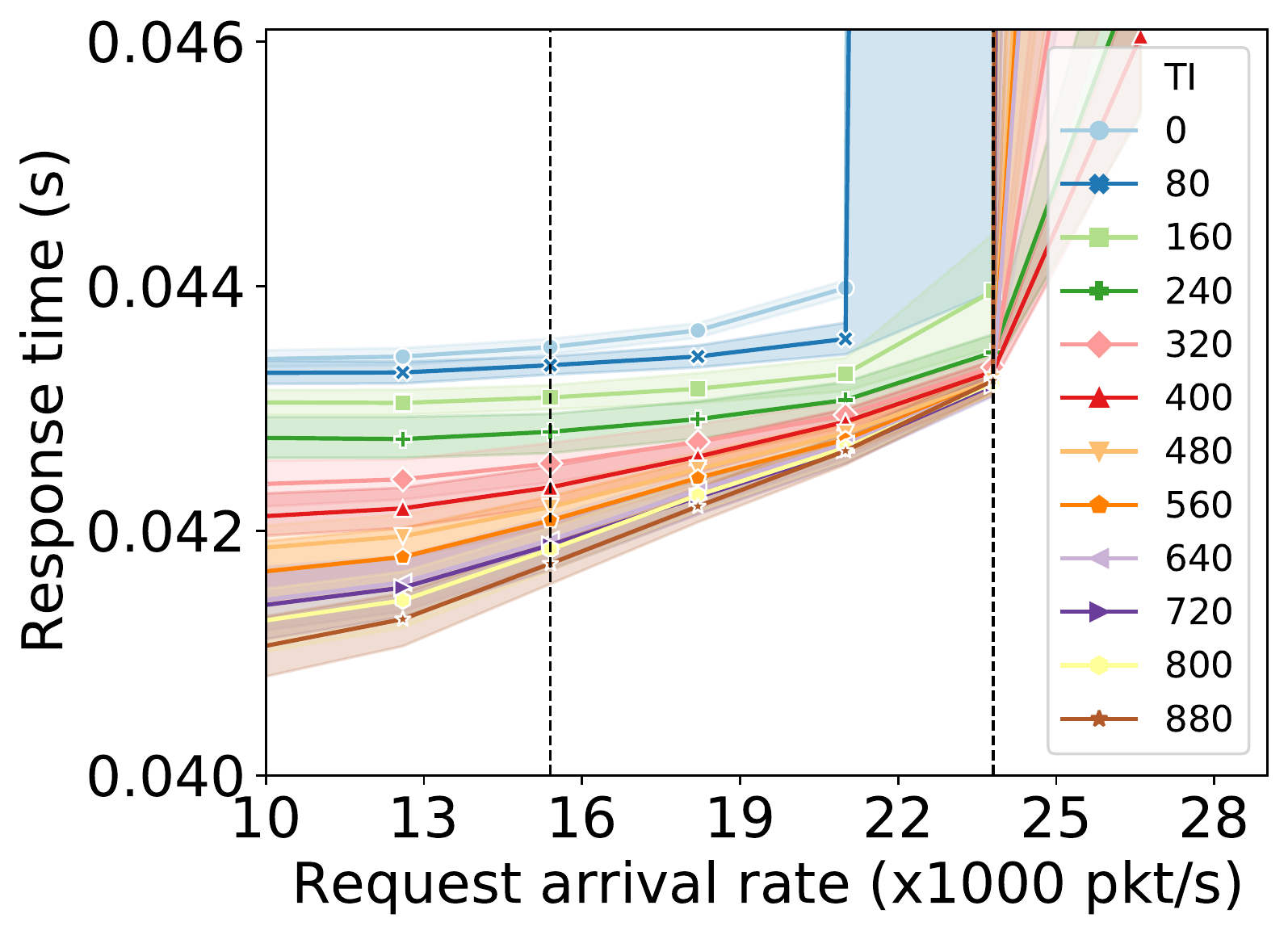}}\\
      \end{center} 
    \caption[Testing performance comparison during the learning process under different request arrival rates in two topologies.]{Testing performance comparison during the learning process under different request arrival rates in two topologies. In particular, ``LWL'' and ``HWL'' indicate the two request arrival rates ($50\%$ and $80\%$ of the total control plane capacity) used during the training.}
    \label{fig:chap5_MAtest_perf_madrl_gamma09}
    \end{figure}

  \textbf{Training effectiveness}: 
  % We first demonstrate the training performance of MA-PPO in two topologies. The learning curves showing the average response time across TIs are demonstrated in \Cref{fig:chap5_MAtrain_perf_sadrl_gamma09}. Similar to SA-PPO, the response time can rapidly converge in both topologies. 
  We investigate how the performance improves as the training proceeds at different TIs. From \Cref{fig:chap5_MAtest_perf_madrl_gamma09}, it can be observed that the policies obtained at the later TIs achieve lower response time compared to those obtained at the earlier TIs, which implies that MA-PPO can effectively improve the performance with continued training of the policy. For example, in Figure~\subref{fig:chap5_southAm_MAtest_gamma09} and Figure~\subref{fig:chap5_southAm_MAtest_gamma09_enlarge}, the response time of the initialized policies (i.e., TI=0) jumps from 90 ms to 2 s when the arrival rate reaches 19k pkt/s. This is mainly because when the policy is randomly initialized, its behaviors are similar to a randomized policy which equally distributes requests among all controllers. Therefore, as the request arrival rate increases, controllers with low capacities are easily overloaded, resulting in high response time. In comparison, the policies obtained after 320 TIs can keep the response time below 1 s under the same request arrival rate. Apart from avoiding overloading controllers at high request arrival rates, the training also consistently reduces the response time when the request arrival rate is low. Similar patterns can also be observed in Asia topology from Figure~\subref{fig:chap5_asia_MAtest_gamma09} and Figure~\subref{fig:chap5_asia_MAtest_gamma09_enlarge}. 
  % We have also included the learning curves showing the average response time across different TIs in Appendix~\ref{sec:MAtrain_perf}.

    \begin{figure}[tb!]
      \begin{center}
        \subfloat[Training]{\label{fig:chap5_southAm_MAtrain_satrain_gamma09}\includegraphics[width=0.47\linewidth]{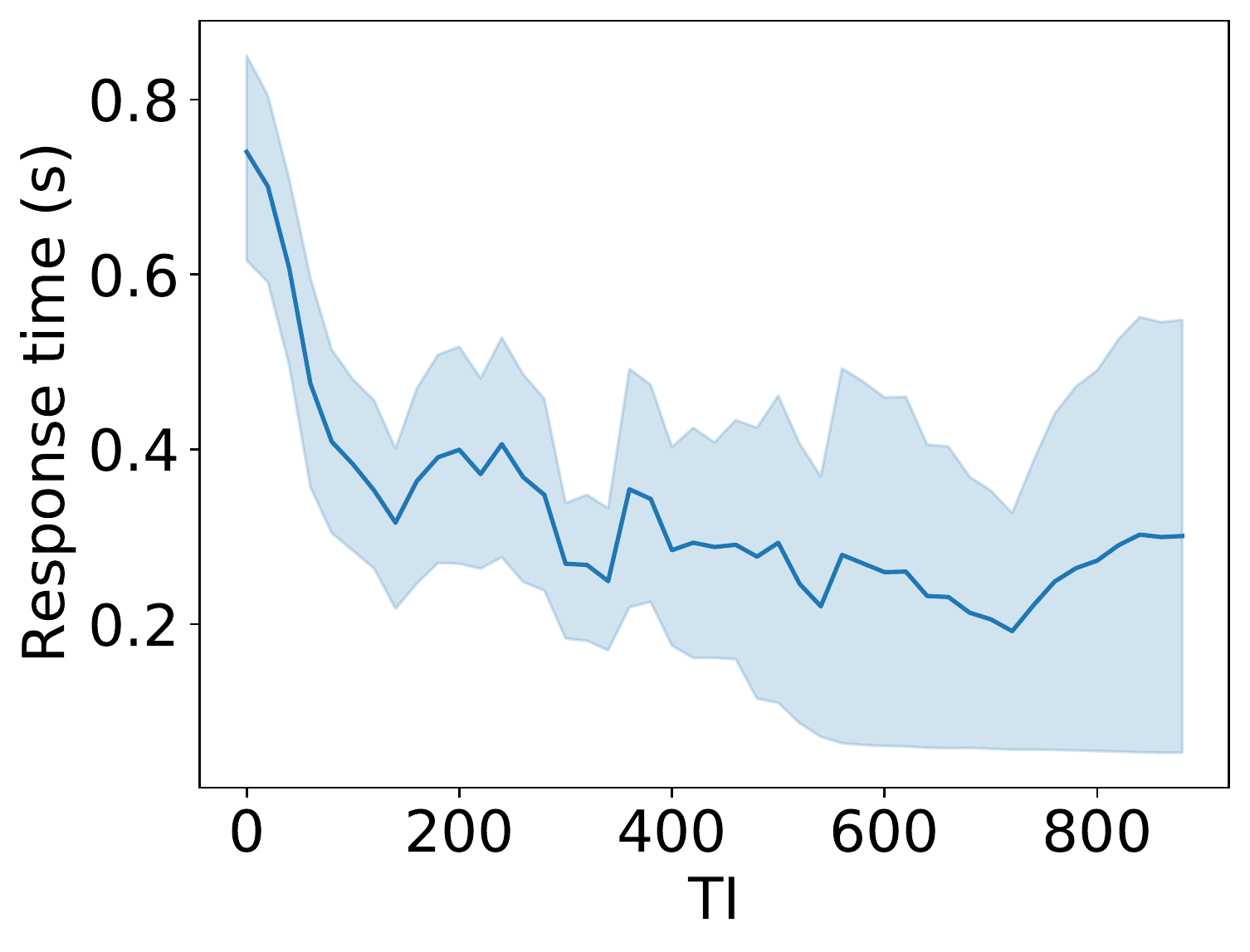}} 
        \subfloat[Testing]{\label{fig:chap5_southAm_MAtest_satrain_gamma09}\includegraphics[width=0.45\linewidth]{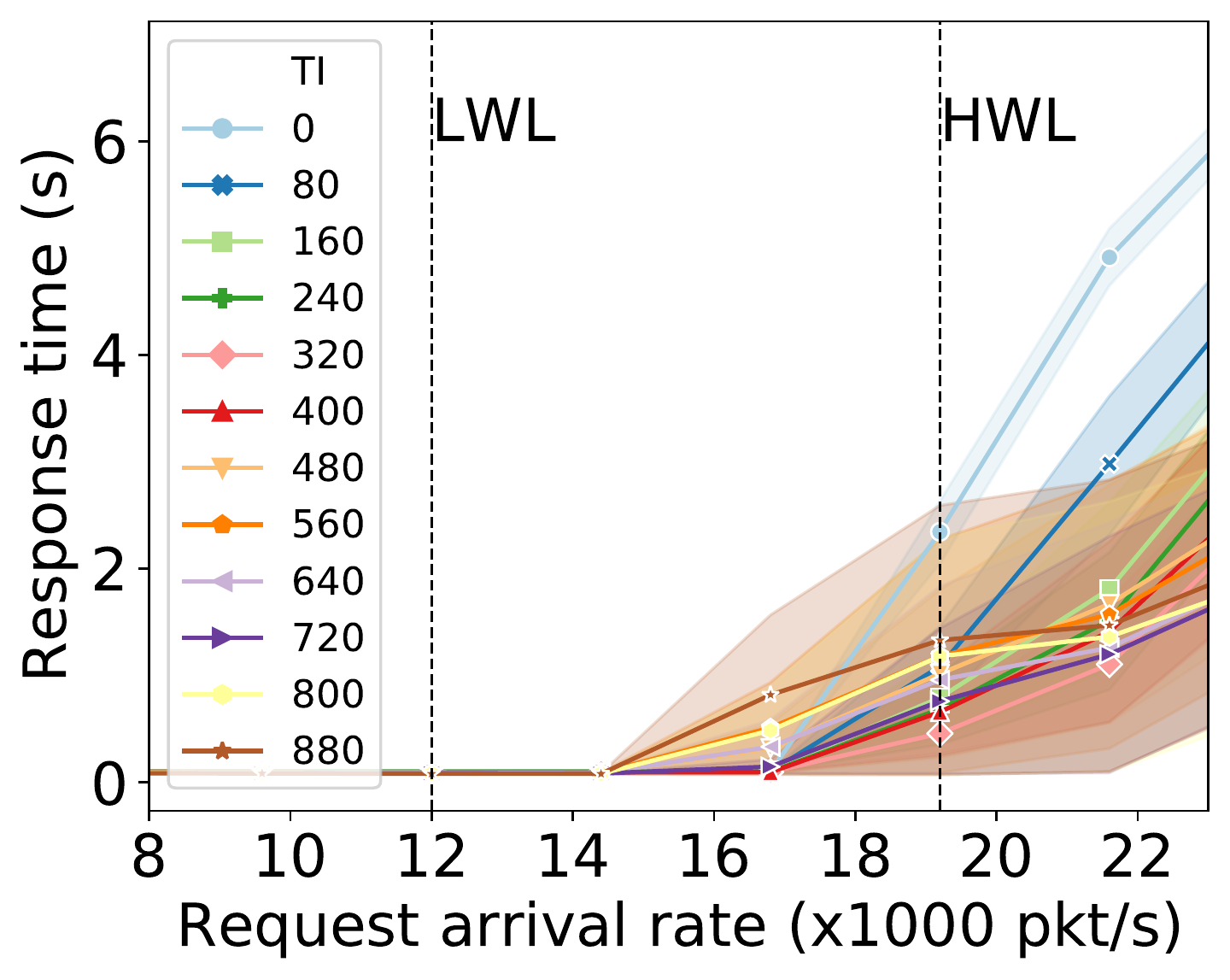}} %\enspace
        % \subfloat[Enlarged bottom area of (a)]{\label{fig:chap5_southAm_MAtest_satrain_gamma09_enlarge}\includegraphics[width=0.5\linewidth]{figures/chapter5/draw_figures/fig/MA_satrain_individual_policy_southAmerican_gamma09_diff_learningStep_testing_Perf_enlarge.pdf}}
      \end{center} 
    \caption{Training and testing performance of SA-PPO-MA in South America Network.}
    \label{fig:chap5_SAtrain_MAtest_perf_madrl_gamma09}
    \end{figure}

    \begin{figure}[!t]
      \begin{center}
        \subfloat[South America]{\label{fig:chap5_southAm_MAtest_ma_vs_satrain_gamma09}\includegraphics[width=0.45\linewidth]{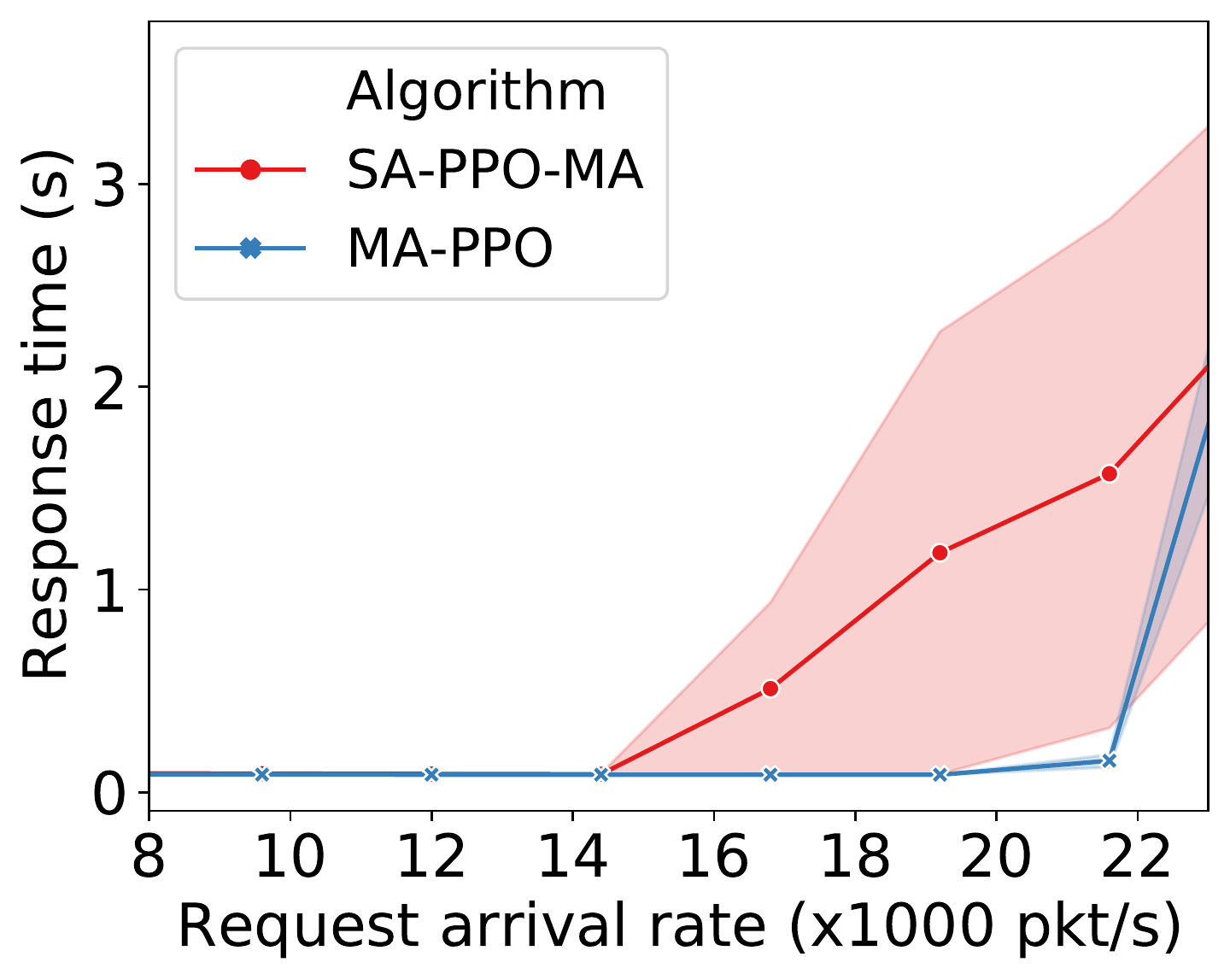}} %\enspace
        % \subfloat[Enlarged bottom area of (a)]{\label{fig:chap5_southAm_MAtest_ma_vs_satrain_gamma09_enlarge}\includegraphics[width=0.5\linewidth]{figures/chapter5/draw_figures/fig/MAvsSA_training_policy_southAm_testing_Perf_enlarge.pdf}}\\
        \;
        \subfloat[Asia]{\label{fig:chap5_asia_MAtest_ma_vs_satrain_gamma09}\includegraphics[width=0.45\linewidth]{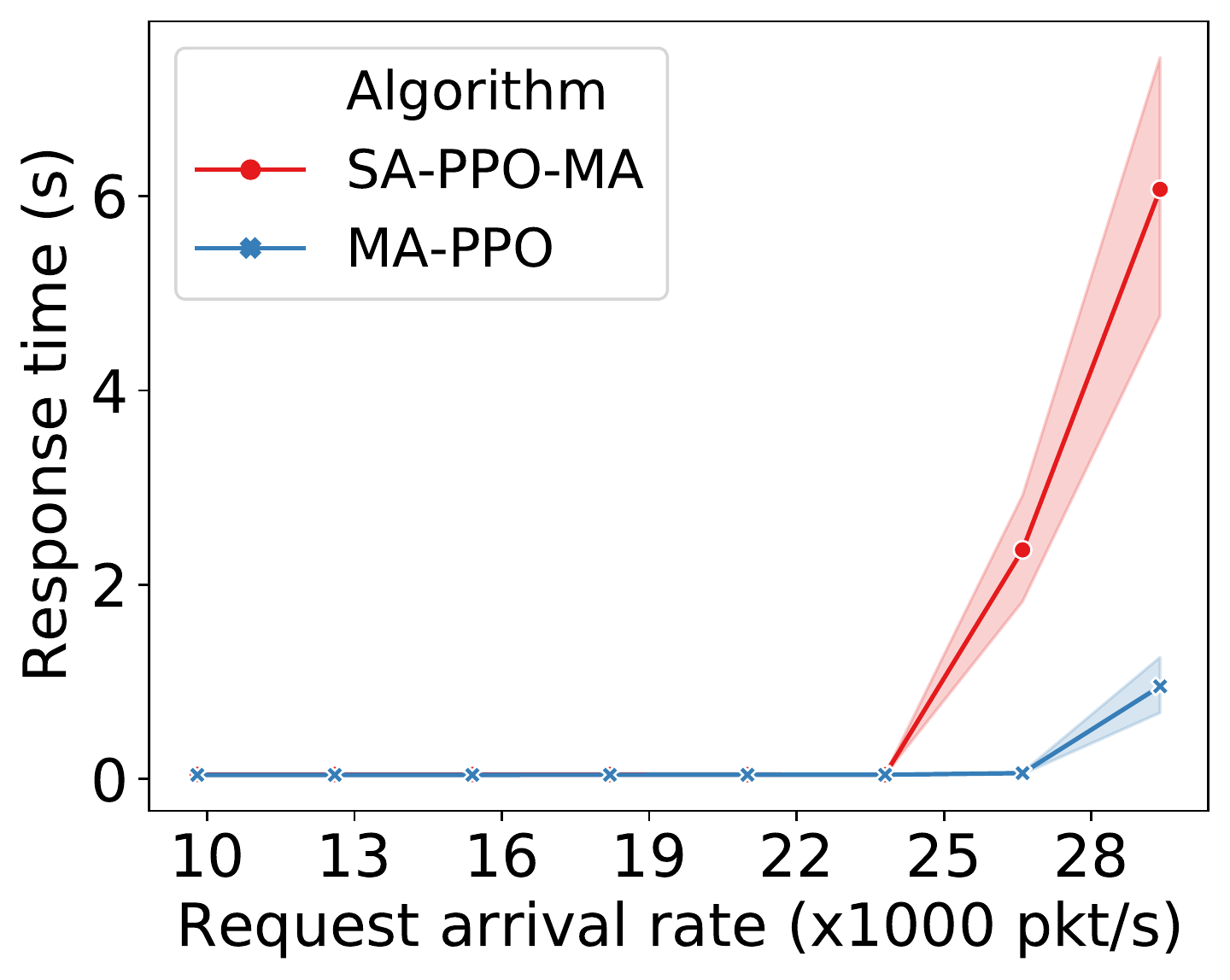}}
        % \subfloat[Enlarged bottom area of (c)]{\label{fig:chap5_asia_MAtest_ma_vs_satrain_gamma09_enlarge}\includegraphics[width=0.5\linewidth]{figures/chapter5/draw_figures/fig/MAvsSA_training_policy_asia_testing_Perf_enlarge.pdf}}
      \end{center} 
    \caption{Performance comparison between MA-PPO and SA-PPO-MA.}
    \label{fig:chap5_MAtest_ma_vs_satrain_perf_madrl_gamma09}
    \end{figure}

  \textbf{Single-agent vs. Multi-agent training in MA-DRL}: To demonstrate the necessity of multi-agent training, we compared training performance between MA-PPO and \mbox{SA-PPO-MA}. As we discussed in \Cref{subsec:ma_ppo}, SA-PPO-MA trains a policy and a value function on each agent independently using single-agent PPO. Both its training and testing performance is shown in \Cref{fig:chap5_SAtrain_MAtest_perf_madrl_gamma09}. 
  During the training process, we can observe a high variance in response time at the later TIs from Figure~\subref{fig:chap5_southAm_MAtrain_satrain_gamma09}, which implies that the learning fails to converge. This observation confirms the non-stationary environment issue when single-agent DRL algorithms are used in a multi-agent environment as we discussed in \Cref{subsec:ma_ppo}. Correspondingly, during the testing process, we can also see from Figure~\subref{fig:chap5_southAm_MAtest_satrain_gamma09} that SA-PPO-MA can keep the response time at a low level when the training arrival rate is low (i.e., the left dotted line LWL). However, it fails to avoid overloading controllers at the high training arrival rate (i.e., the right dotted line HWL). This is mainly because SA-PPO-MA does not consider the impact of other agents during the training. As the request arrival rate increases, the importance of agent cooperation becomes significant and the deficiency of SA-PPO-MA becomes obvious.  

  We also compare the performance of the trained policies via MA-PPO and SA-PPO-MA respectively on two network topologies. \Cref{fig:chap5_MAtest_ma_vs_satrain_perf_madrl_gamma09} confirms that polices trained by MA-PPO can effectively cope with increasing requests through better agent cooperation.

\begin{figure}[!tb]
      \begin{center}
        \subfloat[South America]{\label{fig:chap5_southAm_MAtest_diff_algo_gamma09}\includegraphics[width=0.48\linewidth]{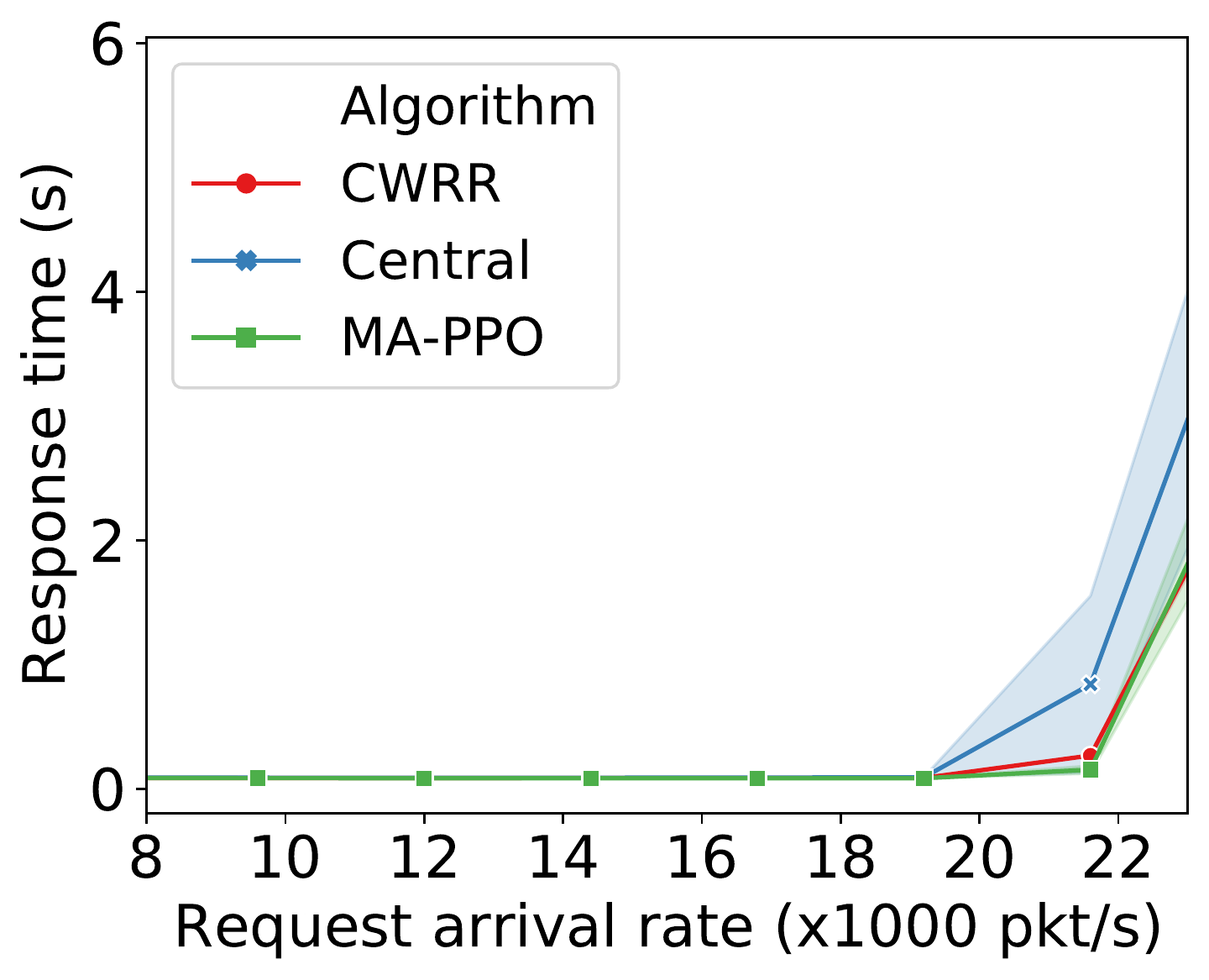}} %\enspace
        \subfloat[Enlarged bottom area of (a)]{\label{fig:chap5_southAm_MAtest_diff_algo_gamma09_enlarge}\includegraphics[width=0.51\linewidth]{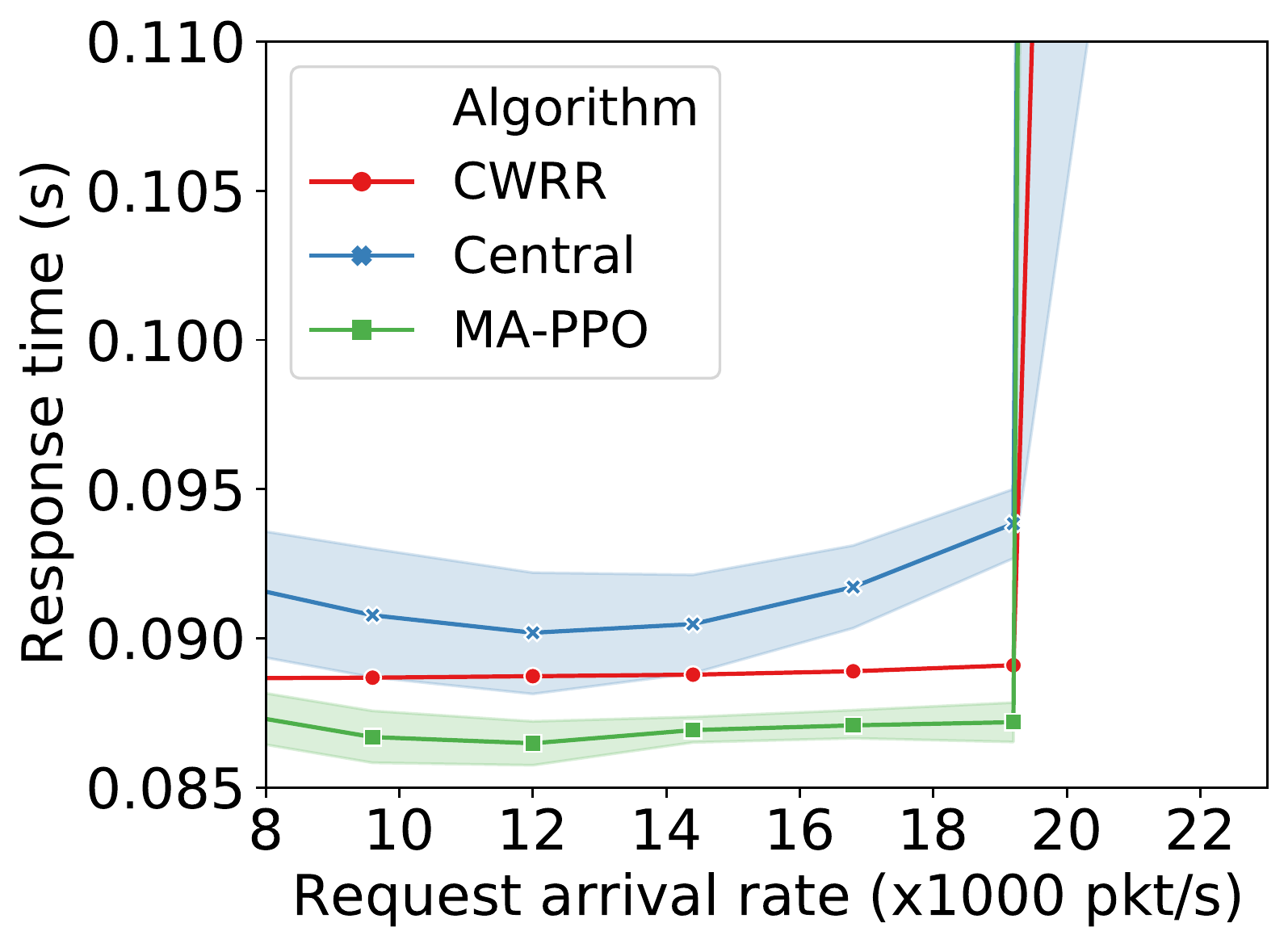}}\\
        \subfloat[Asia]{\label{fig:chap5_asia_MAtest_diff_algo_gamma09}\includegraphics[width=0.48\linewidth]{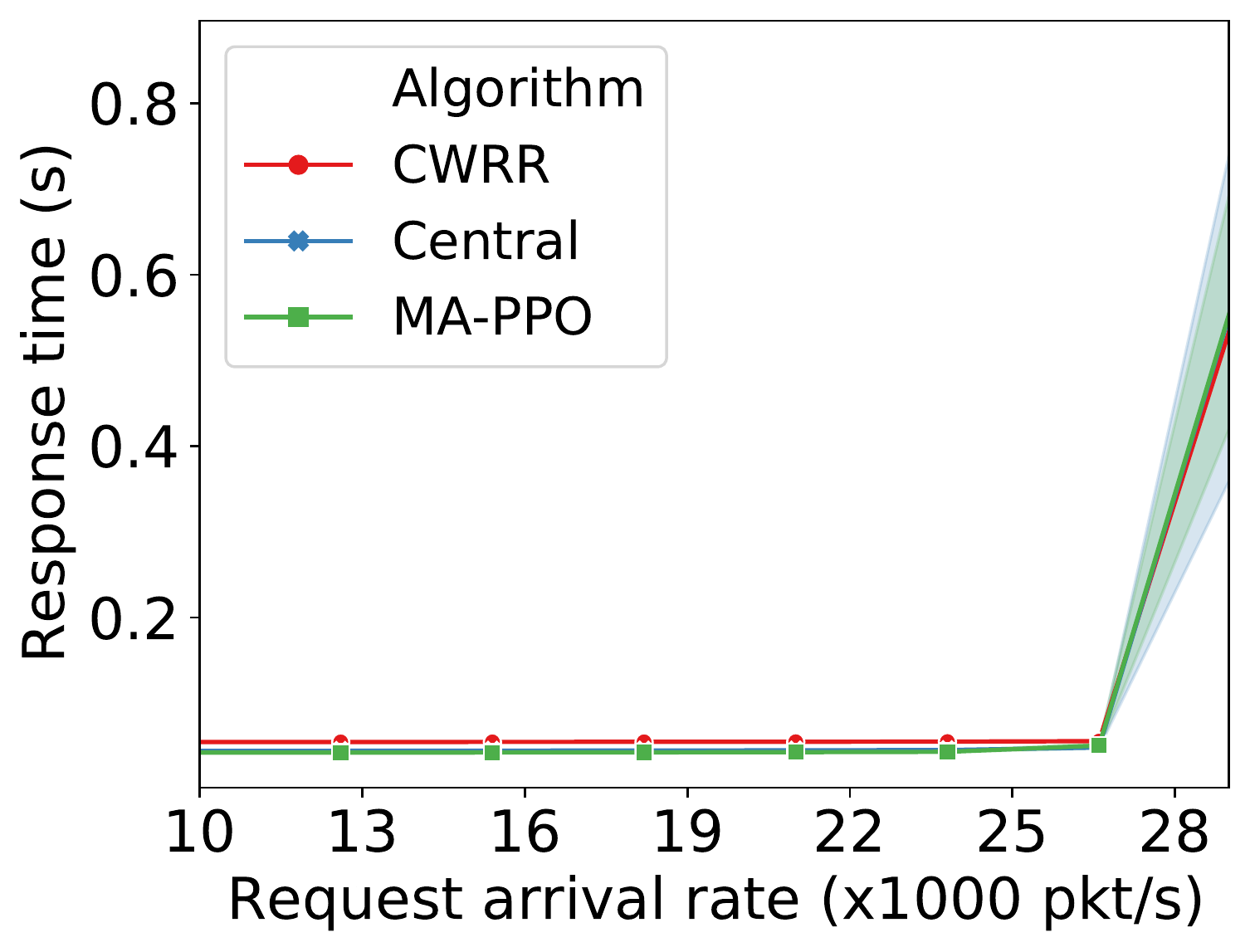}}
        \subfloat[Enlarged bottom area of (c)]{\label{fig:chap5_asia_MAtest_diff_algo_gamma09_enlarge}\includegraphics[width=0.51\linewidth]{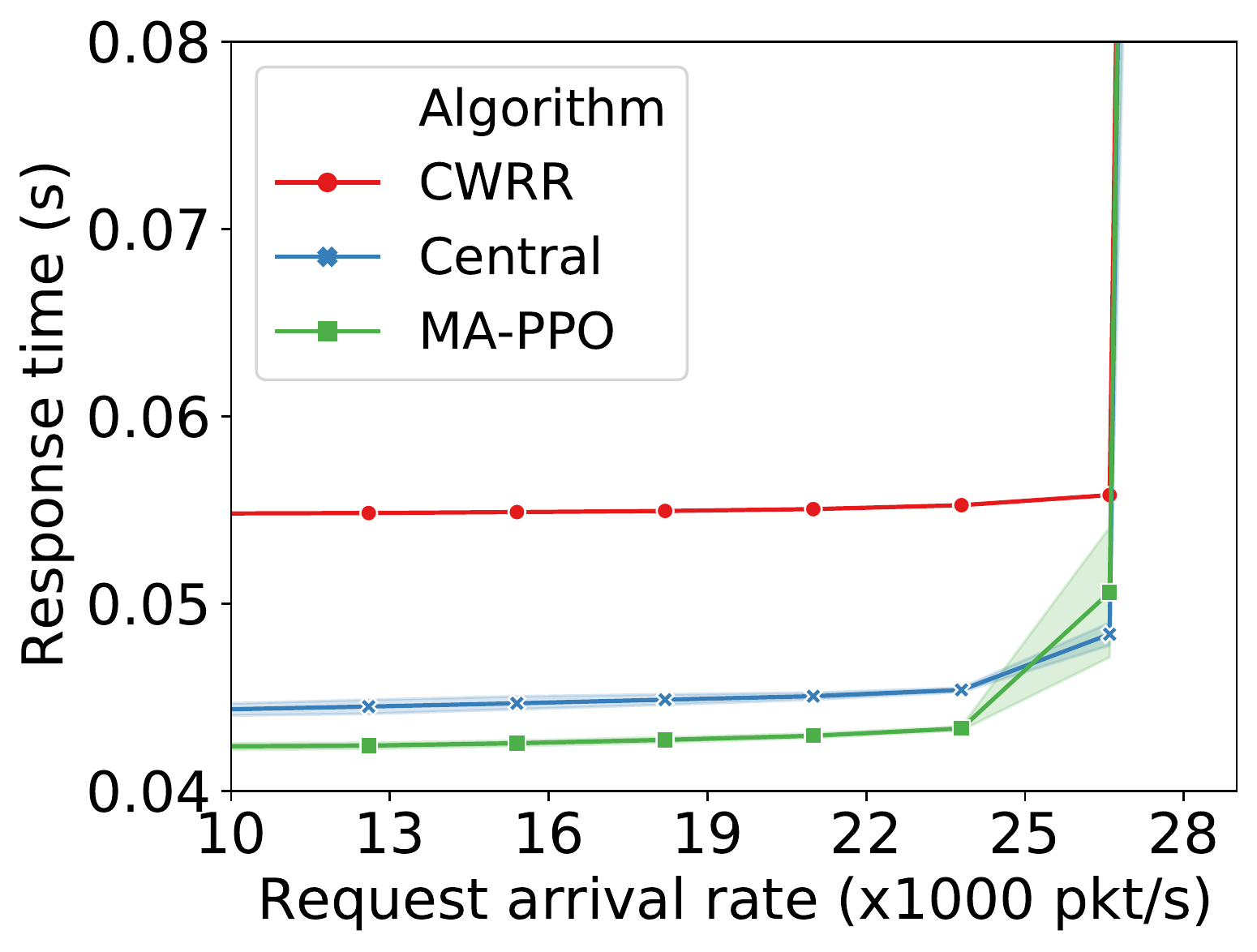}}
      \end{center} 
    \caption{Performance comparison among CWRR, Central, and MA-PPO.}
    \label{fig:chap5_MAtest_diff_algo_gamma09}
  \end{figure}

\begin{figure}[!tb]
      \begin{center}
        \subfloat[South America]{\label{fig:chap5_southAm_MAPPO_vs_GD}\includegraphics[width=0.5\linewidth]{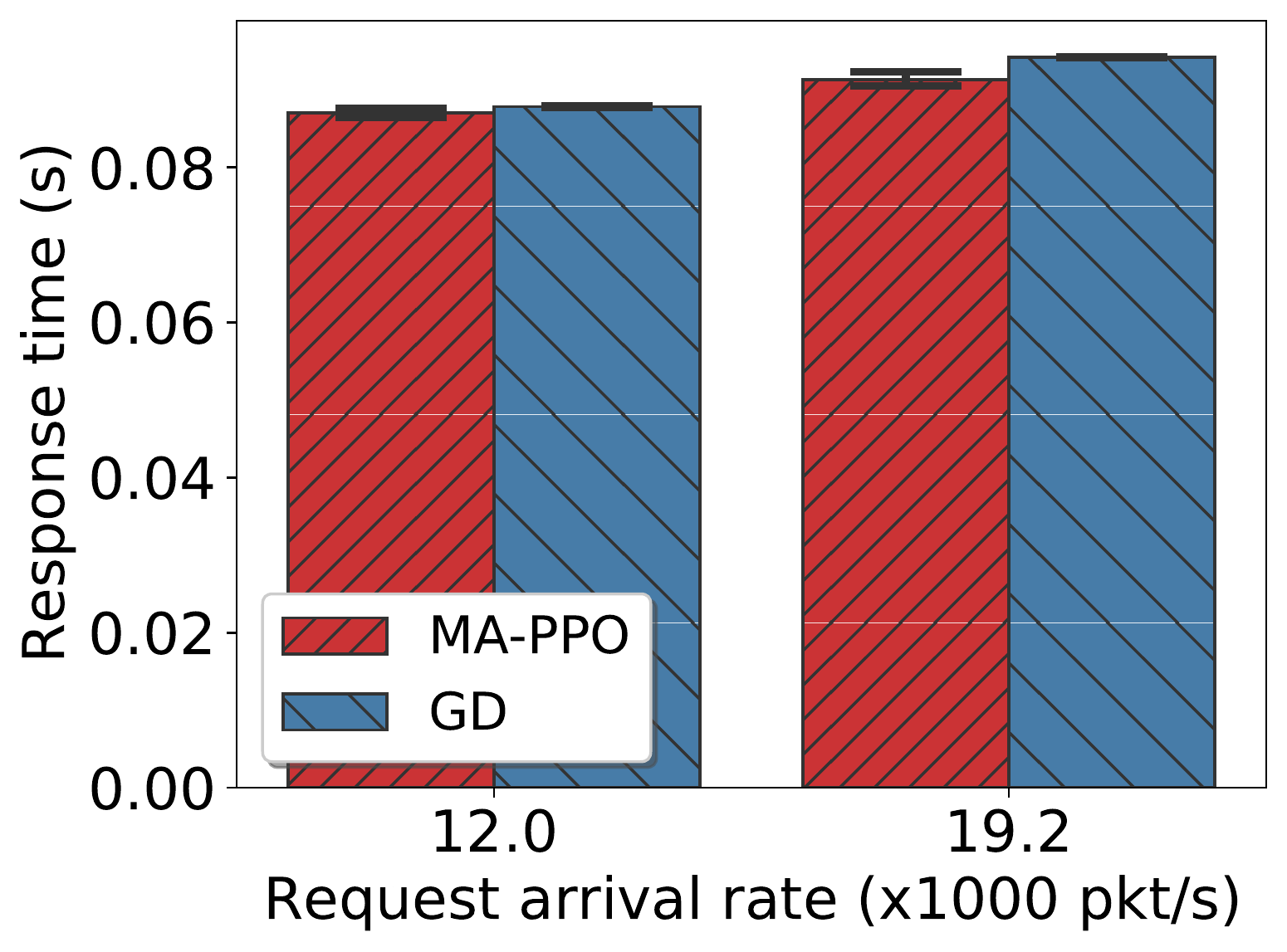}} %\enspace
        \subfloat[Asia]{\label{fig:chap5_asia_MAPPO_vs_GD}\includegraphics[width=0.5\linewidth]{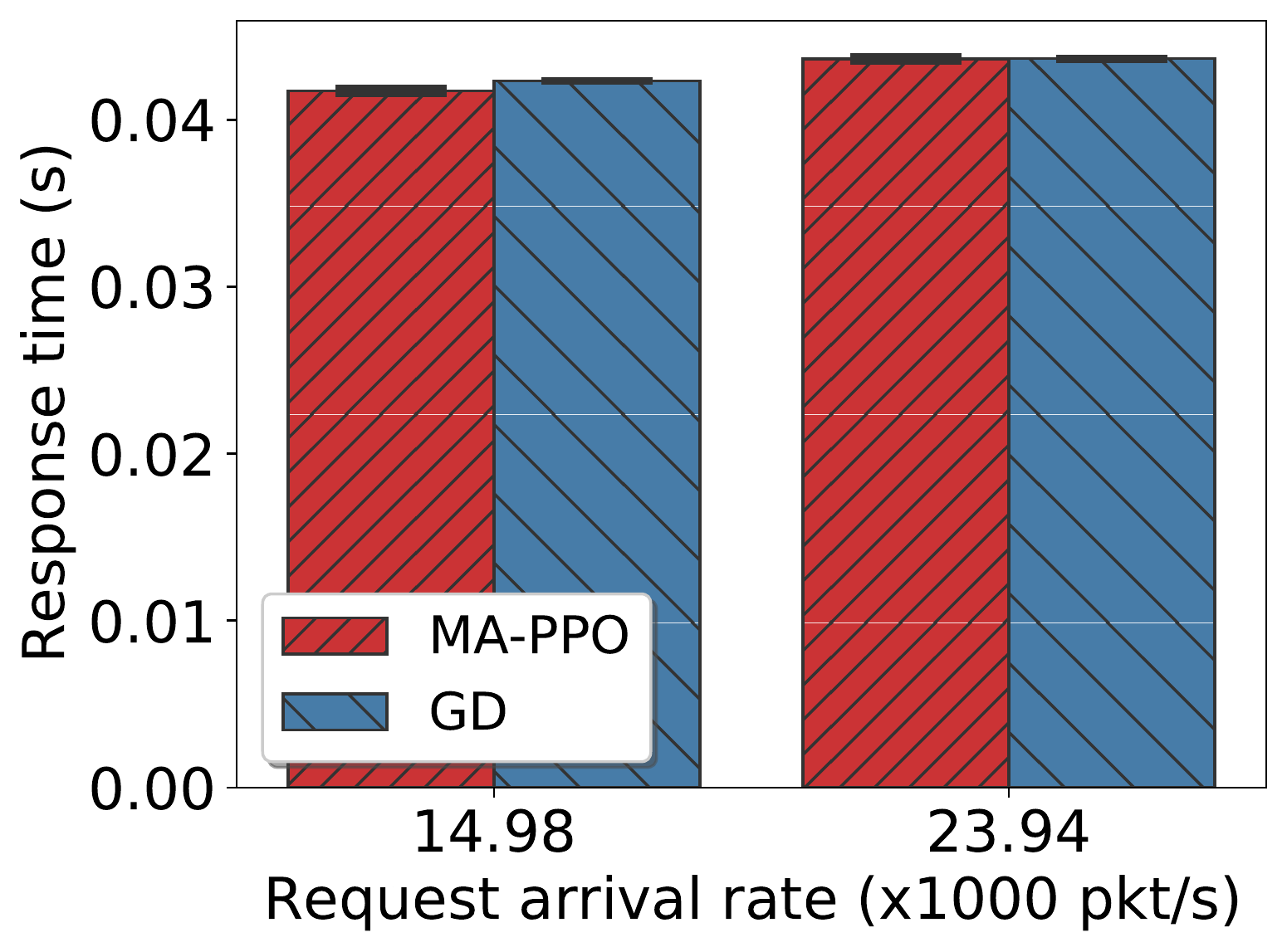}}
      \end{center} 
    \caption{Performance comparison between MA-PPO and GD.}
    \label{fig:chap5_MAPPO_vs_GD}
  \end{figure}

  \textbf{Performance comparison with existing policies}: We compare MA-PPO with several policies
  % as summarized in Appendix~\ref{sec:comparedAlg}
  : (1) a widely used man-made policy (CWRR), a GD-based policy (GD), and the centralized single-agent policy (Central). Results are shown in \Cref{fig:chap5_MAtest_diff_algo_gamma09,fig:chap5_MAPPO_vs_GD}.

  In particular, we can see that in both topologies (Figure~\subref{fig:chap5_southAm_MAtest_diff_algo_gamma09_enlarge} and Figure~\subref{fig:chap5_asia_MAtest_diff_algo_gamma09_enlarge}), the response time of CWRR remains stable because the number of requests dispatched to each controller is proportional to its capacity, which effectively prevents overloading any controller at an early stage. However, solely sending requests based on the controller capacity may not achieve the optimal network performance. Especially, when the workload of the control plane is low, dispatching more requests to a closer controller without overloading it is a better option. In DRL, the relationship between the network performance and RD probabilities is learned during the interaction between the agents and the environment. Therefore, we can see from both Figure~\subref{fig:chap5_southAm_MAtest_diff_algo_gamma09_enlarge} and Figure~\subref{fig:chap5_asia_MAtest_diff_algo_gamma09_enlarge} that MA-PPO achieves a lower response time compared to CWRR. Apart from that, we also notice that MA-PPO achieves a lower response time than Central, which confirms that using a centralized agent can introduce additional propagation latencies.

  MA-PPO is also compared with GD, a model-based optimization approach. As shown in \Cref{fig:chap5_MAPPO_vs_GD}, MA-PPO can achieve slightly lower response time. This is mainly because GD optimizes the response time for given network information. However, in practice, network information such as request arrival rates can only be estimated. The inaccurate network information hinders GD achieving its optimal performance. 

  Even though both GD and MA-PPO achieve similar performance, MA-PPO has the advantage of low computation and communication overheads. During our simulation, we notice that the execution time of GD is 10 times longer than running the MA-PPO policy. The reason is that to obtain the request dispatching probabilities for the entire data plane (i.e., all switch centers), GD needs to iteratively perform gradient calculation. This can be computational intensive especially for a network with a large number of switch centers. In comparison, each agent in MA-PPO runs its policy individually (a forward pass from NN input layer to output layer) to calculate the controller priorities used by the switch center where the agent is placed on. Apart from that, for communication overheads, GD requires the information collected from the entire network while MA-PPO only uses its local network observation. Therefore, MA-PPO is more suitable for large-scale networks.

\section{Conclusion}
In this paper, we achieved the research goal of optimizing network performance by designing RD policies to properly utilize controller resources. For this purpose, an MA-DRL approach was proposed for learning adaptive RD policies for SDN switches. In particular, the RD policy design problem was formulated as an MA-MDP. To allow our policy to adapt easily to a varying number of controllers, a new adaptive design was proposed to support DNN-based policies. In line with the new policy design, MA-PPO was developed to enable the adaptive policy training by using a new policy gradient calculation technique. 

To demonstrate the effectiveness of our adaptive policy design, our policy was compared with a non-adaptive policy which is widely adopted in the literature. The results showed that the adaptive policy can reduce the DNN complexity without performance degradation. Apart from that, the new policy design performed consistently well with a changing number of controllers. 
Moreover, to evaluate the effectiveness of MA-PPO, extensive simulation studies were conducted which showed that the policy trained via MA-PPO significantly outperformed man-made policies, model-based policies, as well as the policy learned via single-agent DRL. Our approach not only addresses the RD problem in SDN, it can also be applied to facilitate the operations of T-SAC, an anycast-based CDN architecture proposed in~\cite{fu2018taming}.

\ifCLASSOPTIONcaptionsoff
  \newpage
\fi

% \end{thebibliography}
\bibliographystyle{IEEEtran}
\bibliography{JSAC_MADRL}

\vfill


\end{document}